\documentclass[fleqn,usenatbib,useAMS]{mnras}
\usepackage{graphicx}
\graphicspath{{Figures/}}
\usepackage{import}
\usepackage{physics}
\usepackage{mathtools}
\usepackage{relsize}  
\usepackage[T1]{fontenc}
\usepackage{ae,aecompl}
\usepackage{newtxtext,newtxmath}

\newcommand{\HI}{\ion{H}{I}}

\title[EoR with CDM and WDM]{Impact of dark matter models on the EoR 21-cm signal bispectrum}

\author[Saxena et al.]{Anchal Saxena,$^{1}$\thanks{E-mail: msc1803121001@iiti.ac.in (AS)}
Suman Majumdar,$^{1}$
Mohd Kamran$^{1}$ and 
Matteo Viel$^{2,3,4,5}$
\\
$^{1}$Discipline of Astronomy, Astrophysics and Space Engineering, Indian Institute of Technology Indore, Simrol, Indore 453552, India\\
$^{2}$SISSA, Via Bonomea 265, 34136 Trieste, Italy\\
$^{3}$IFPU, Via Beirut 2, 34014 Trieste, Italy\\
$^{4}$INFN-Sezione di Trieste, via Valerio 2, 34127 Trieste, Italy\\
$^{5}$INAF-OATS, via Tiepolo 11, 34131 Trieste, Italy\\
}

\date{Accepted XXX. Received YYY; in original form ZZZ}

\pubyear{2020}

\begin{document}
\label{firstpage}
\pagerange{\pageref{firstpage}--\pageref{lastpage}}
\maketitle

\begin{abstract}
The nature of dark matter sets the timeline for the formation of first collapsed haloes and thus affects the sources of reionization. Here, we consider two different models of dark matter: cold dark matter (CDM) and thermal warm dark matter (WDM), and study how they impact the epoch of reionization (EoR) and its 21-cm observables. Using a suite of simulations, we find that in WDM scenarios, the structure formation on small scales gets suppressed, resulting in a smaller number of low mass dark matter haloes compared to the CDM scenario. Assuming that the efficiency of sources in producing ionizing photons remains the same, this leads to a lower number of total ionizing photons produced at any given cosmic time, thus causing a delay in the reionization process. We also find visual differences in the neutral hydrogen ($\HI$) topology and in 21-cm maps in case of the WDM compared to the CDM. However, differences in the 21-cm power spectra, at the same neutral fraction, are found to be small. Thus, we focus on the non-Gaussianity in the EoR 21-cm signal, quantified through its bispectrum. We find that the 21-cm bispectra (driven by the $\HI$ topology) are significantly different in WDM models compared to the CDM, even for the same mass averaged neutral fractions. This establishes that the 21-cm bispectrum is a unique and promising way to differentiate between dark matter models, and can be used to constrain the nature of the dark matter in the future EoR observations.
\end{abstract}


\section{Introduction}
\label{sec:intro}

Cosmology is the study of the evolutionary history of the universe, and one crucial missing chapter in this history is the Cosmic Dawn and Epoch of Reionization. This was the period when the first sources of light in the universe were formed, and these and subsequent population of sources emitted the high energy X-ray and UV radiation, which in turn heated up and reionized the intergalactic medium  \citep[see][for reviews]{2001PhR...349..125B, 2006PhR...433..181F}. This is the period in cosmic history when the universe has witnessed the formation of the first bound structures. Thus, the CD-EoR has significant implications on the large scale structures that we see around us today.

After the cosmological recombination, the universe went into the dark ages during which the density fluctuations in the matter distribution grew, and after reaching a threshold, the matter collapsed to make the first bound objects. The nature of dark matter sets the timeline and characteristics of these first bound objects, which were the hosts for the first sources of light, so it is essential to see the impact of different dark matter models on the observables from the Cosmic Dawn and EoR. Then, the natural question that arises is whether one can use the differences in these observables, estimated for different dark matter models, in order to constrain the nature of dark matter. The present observational probes that allow us to have a peak in this epoch are the absorption spectra of high redshift quasars \citep{2001ARA&A..39...19L, 2003AJ....126....1W,boera19} and the Thomson scattering optical depth of the Cosmic Microwave Background Radiation photons \citep{2003ApJ...583...24K, 2011ApJS..192...18K}. However, these indirect probes provide very limited and weak constrains on the CD-EoR. The $\HI$ 21-cm line, which arises due to the hyperfine splitting of the ground state of the neutral hydrogen, is a direct and most promising probe to study this period. Motivated by this, a large number of radio interferometers, including the GMRT \citep{2013MNRAS.433..639P}, LOFAR \citep{mertens20, ghara20}, MWA
\citep{barry19, li19}, and PAPER \citep{kolopanis19} are attempting a statistical detection of this signal using the power spectrum statistic. In parallel, there is a complementary approach to detect the sky averaged global 21-cm signal from the CD-EoR using experiments e.g. the EDGES \citep{2018Natur.555...67B}, DARE \citep{2017ApJ...844...33B}, and SARAS \citep{2018ApJ...858...54S}. The next-generation interferometers like the SKA \citep{2015aska.confE...1K, mellema2015hi} are expected to see a giant leap in the sensitivity, which will enable them to make tomographic images of the $\HI$ distribution across cosmic time.

The nature of the dark matter is mostly unknown to us. We can classify the dark matter into cold, warm, and hot categories based on the free streaming length scale of the dark matter particles \citep[see][and references therein]{zora75587}. The $\Lambda$CDM cosmology is consistent with several observations at large scales, including the observations of Lyman-$\alpha$ forest at small and medium scales, clusters, and CMB anisotropies. However, some discrepancies between the theory and observations arise at small scales $\leq 1.0$ Mpc. Some of these are labelled as the too-big-to-fail problem, the core-cusp problem, satellite abundance, and galaxy abundance in mini-voids \citep[see][for more details]{2009NJPh...11j5029P,2015PNAS..11212249W, 2017ARA&A..55..343B}. These issues may be resolved either by invoking astrophysical baryonic processes or by assuming the dark matter to be warm instead of cold. Cosmology and especially structure formation and evolution can thereby be a good probe for studying such dark sectors \citep{zora75587, 2013PhRvD..88d3502V}.

Two potential candidates for the warm dark matter are sterile neutrinos and gravitinos, both of which require the extensions of the standard model of particle physics \citep{1994PhRvL..72...17D, 2005PhRvD..71f3534V, 2019PrPNP.104....1B}. However, in this work, we consider a thermal relic to be the candidate for the WDM. Unlike sterile neutrinos, this candidate is probably
less motivated but easier to simulate given the fact that its transfer function has been more studied by several authors, also in terms of non-linear structure formation.
Several studies have constrained the WDM particle mass using galaxy luminosity function of high redshift galaxies and Lyman-$\alpha$ forest data \citep{10.1093/mnras/stu719, PhysRevD.95.083512, 2017ApJ...836...16D, Safarzadeh_2018} and found a robust lower bound on the thermal warm dark matter particle mass to be $\geq$ 2 keV. 
Recent Lyman-$\alpha$ forest power spectrum analyses point to somewhat tighter
limits $>3.5$ keV at the 2$\sigma$ C.L. \citep{irsic17}. However, for the purposes of the present work, it is appropriate to investigate the values of the thermal masses that are also in principle already ruled out by other observables.

In this work, we consider the standard CDM model and thermal warm dark matter (WDM) with masses of the thermal relic of 2 and 3 keV. Investigations of the structure formation processes, in the WDM scenario and especially at high redshifts, have been performed by e.g.\,\citep{maio15} (and references therein).
Recently, reionization has been studied in the warm dark matter models or extensions of the standard CDM scenario by several authors \citep{gao07, lovell14, 2014MNRAS.438.2664S, 2017ApJ...836...16D, 2018ApJ...852..139V, 2018JCAP...08..045D, lapidanese15, carucci15,carucci17}. These studies have focused on the 21-cm power spectrum and found the differences in the 21-cm power spectrum to be significantly small between different models of the dark matter, and probably not large enough to be detectable by the first generation of radio interferometers. 

However, the 21-cm power spectrum can provide a complete description of a signal only if it is a Gaussian random process in nature. Whereas, the EoR 21-cm signal is highly non-Gaussian, specifically during the intermediate and later stages of reionization (see also \citep{pillepich07}). The power spectrum, therefore, can not capture this non-Gaussian feature of the signal. The effort to differentiate between different models of dark matters using the 21-cm power spectrum will not be optimal as it does not capture the non-Gaussian information present in the signal. To capture this evolving non-Gaussianity, one would need to use a higher-order statistic, the bispectrum, which is the Fourier equivalent of the 3-point correlation function. Recently, the CD-EoR 21-bispectrum has been investigated using both analytical models and numerical simulations \citep{2005MNRAS.358..968B, 10.1093/mnras/stw482, 2018MNRAS.476.4007M, 10.1093/mnras/sty2740, 2020MNRAS.492..653H}.
These authors have independently confirmed that the two major sources of non-Gaussianity in this signal are the matter density fluctuations and the neutral fraction fluctuations. In this work,  we aim to quantify the differences in the observables of the CD-EoR 21-cm signal for different dark matter models. Through this analysis, we would like to identify the optimal statistics that can be used to distinguish between different dark matter models and their signature on the CD-EoR 21-cm signal.  

This article is organized as follows: In section~\ref{stru}, we briefly discuss the formation of structures in different dark matter models. In section~\ref{method}, we concisely describe our semi-numerical approach to simulate the redshifted 21-cm signal and methods to estimate different statistics out of it. Section~\ref{results} describes our results. In section~\ref{summary}, we summarize our findings.  

Throughout this work, we have used the values of cosmological parameters as $\Omega_{\rm m} = 0.308$, $\Omega_{\rm b} = 0.048$, $\Omega_\Lambda = 0.692$, $h = 0.678$ and $\sigma_8 = 0.829$ \citep{2016A&A...594A..13P}.

\section{Simulating the structure formation for different dark matter models} \label{stru}
The dark matter can be classified into three categories, based on their characteristics; they are - cold, warm, and hot. This classification is done according to the free streaming scale of the dark matter particles \citep{2001ApJ...556...93B, 2001PhRvD..64b3501A,2013MNRAS.433.1573S}. In the early Universe, primordial density perturbations on the scales smaller than the free streaming length scale get damped. This is because the dark matter particles at these scales stream out from the overdense to the underdense regions, whereas the fluctuations on the scales larger than this scale remain unaffected.
\begin{figure*}
    \centering
    \includegraphics[width = \textwidth]{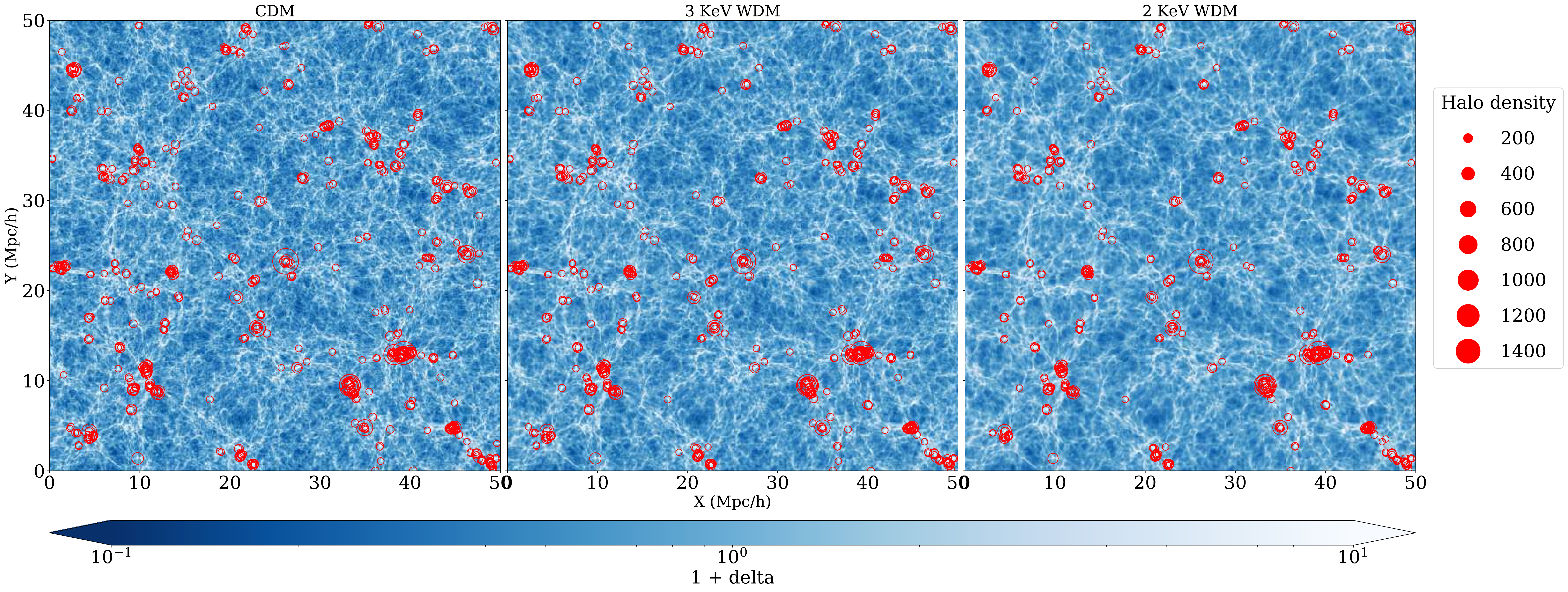}
    \caption{Dark matter overdensity field obtained from our simulation, over-plotted with the halo mass field at z = 8 using CIC algorithm for CDM (left-hand panel), 3 keV WDM (middle panel) and 2 keV WDM model (right-hand panel). The color-bar represents the dark matter overdensity 1 + $\delta$, and the red circles represent the halo mass density with their radius proportional to the halo mass estimated at that grid. Halo mass density is represented in the internal grid units of the simulation.}
    \label{fig:odf}
\end{figure*}

The free streaming scale can be defined as the length traversed by a dark matter particle before the density perturbations start to grow significantly. It can be calculated as \citep{1990eaun.book.....K,2011PhRvD..84f3507S}
\begin{equation} \label{lam1}
    \lambda_{{\rm fs}} = \int_{0}^{t_{{\rm mre}}} \frac{v(t)\,{\rm d}t}{a(t)} \approx \int_{0}^{t_{{\rm nr}}} \frac{c\,{\rm d}t}{a(t)} + \int_{t_{{\rm nr}}}^{t_{{\rm mre}}} \frac{v(t)\,{\rm d}t}{a(t)}\,,
\end{equation}
where $t_{{\rm mre}}$ is the epoch of matter-radiation equality, and $t_{{\rm nr}}$ represents the onset of non-relativistic behavior of dark matter particles. Also, we have made use of the fact that in relativistic domain, $v(t) \sim c$. In the non-relativistic regime as $v(t) \sim a(t)^{-1}$ and during the radiation dominated era, $a(t) \propto t^{1/2}$, so equation~(\ref{lam1}) leads to
\begin{equation}
    \lambda_{{\rm fs}} \sim \frac{2 c t_{{\rm nr}}}{a_{{\rm nr}}} \left[ 1 + \log \left( \frac{a_{{\rm mre}}}{a_{{\rm nr}}}  \right) \right]\,.
\end{equation}
So, by increasing the time when the nature of dark matter particle became relativistic i.e.$\,t_{{\rm nr}}$, we can make the free streaming scale $\lambda_{{\rm fs}}$ larger. We can calculate the mass of a halo whose formation is suppressed due to the free streaming as
\begin{equation} \label{eq:M_FS}
    M_{{\rm fs}} = \frac{4}{3} \upi \left(  \frac{\lambda_{{\rm fs}}}{2} \right)^3 \bar{\rho}\,.
\end{equation}

In the cold dark matter scenario, the dark matter particles become non-relativistic already at the time of decoupling $t_{{\rm nr}}$ $\sim$ $t_{{\rm dec}}$, leading to a negligible free streaming length.  Hence, it does not erase the initial perturbations on small scales, and we have a bottom-up scheme for the formation of structures in the CDM cosmology, with galaxies forming first and galaxy clusters later. Several dark matter simulations confirm this \citep{2011ASL.....4..297D, 2012AnP...524..507F}. However, this model is inconsistent with the observations on small scales.

In the warm dark matter scenario, the particles become non-relativistic later compared to the CDM model and the $t_{{\rm nr}}$ lies somewhere in between $t_{{\rm dec}}<t_{{\rm nr}}<t_{{\rm mre}}$. Thus, the free streaming scale of the dark matter particles in the WDM model is larger compared to the CDM model leading to suppressed density perturbations at small scales. It results in a bottom-up structure formation at scales greater than $\lambda_{{\rm fs}}$ and a top-down structure formation at scales less than $\lambda_{{\rm fs}}$. Also, various simulations of warm dark matter models confirm this \citep{2001ApJ...556...93B, 2012MNRAS.424..684S, 2013PhRvD..88d3502V}. 

\begin{figure}
    \centering
    \includegraphics[width = \columnwidth]{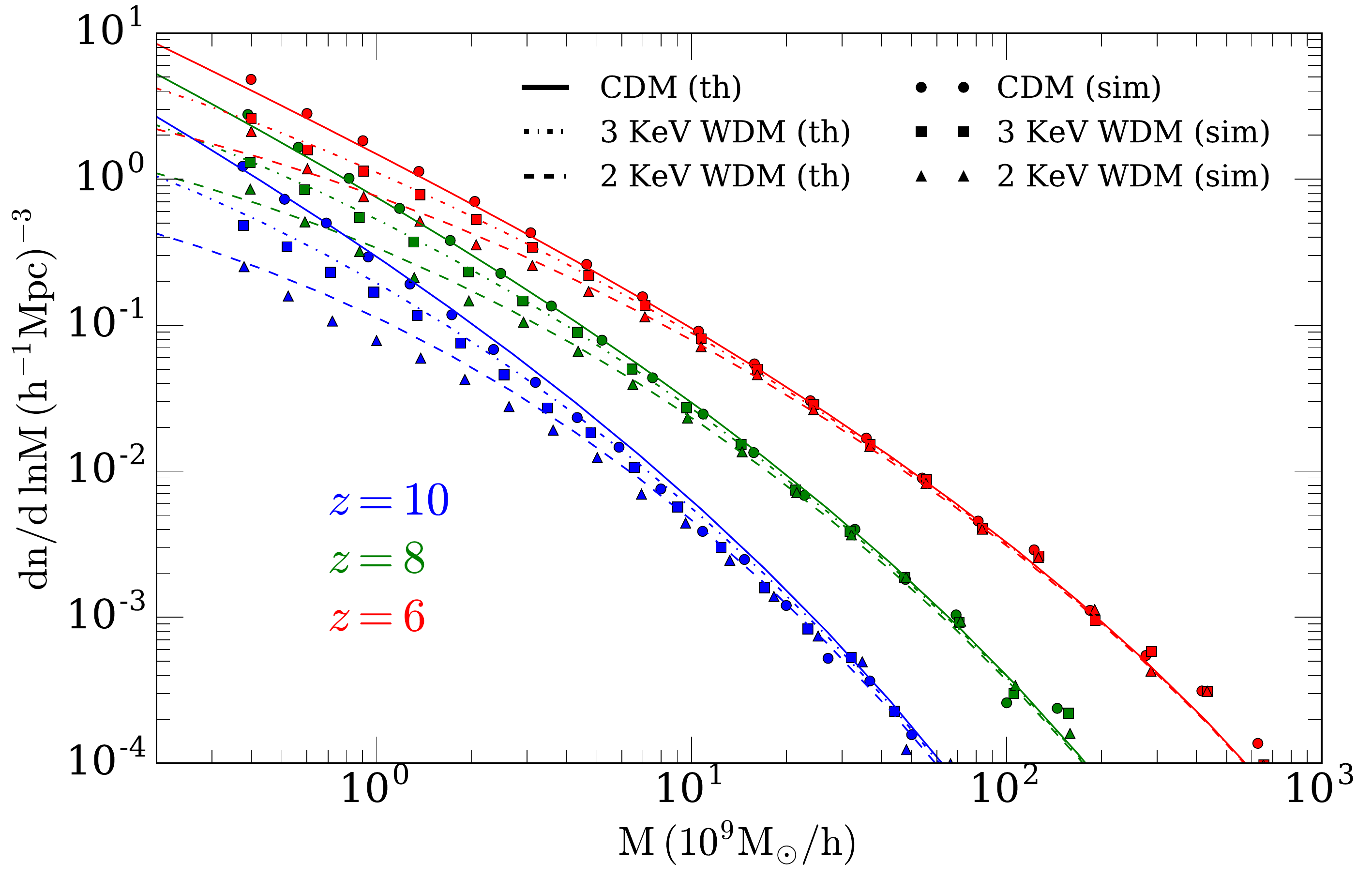}
    \caption{Halo mass function obtained through simulations for CDM (circles), 3 keV WDM (squares) and 2 keV WDM (triangles) model at $z$ = 10 (blue), 8 (green) and 6 (red). The Sheth-Tormen mass function estimated using the top-hat window function is shown for CDM (solid), 3 keV WDM (dotted-dashed) and 2 keV WDM (dashed) model.}
    \label{fig:hmf}
\end{figure}

We used GADGET 2.0 N-body simulations \citep{2005MNRAS.364.1105S} for dark matter for a volume of 50 h$^{-1}$ com. Mpc$^3$ with $1024^3$ dark matter particles. The gravitational softening is chosen
to be 1/30 of the mean linear interparticle separation. The mass of each DM
particle is $9.95\times 10^6$ M$_{\odot}/h$, thereby DM haloes of 10$^9$ M$_{\odot}/h$ are resolved with about 100 particles.
Compared to other investigations of DM structure formation, this is more focussed on the small scales, and the volume and resolution used here are ideal for probing the 21cm signal and the WDM induced effects.
Initial conditions for WDM are generated according to the fitting formula presented in \citep{2005PhRvD..71f3534V}, with thermal velocities drawn from a Fermi-Dirac distribution.
The simulations' snapshots store the position and velocities of the dark matter particles at these predesignated redshifts. Next, we use a Friends-of-Friend algorithm to identify the collapsed gravitationally bound objects (haloes) in these dark matter fields. 

Figure~\ref{fig:odf} shows the two-dimensional slices of the matter overdensity field over-plotted with the halo fields obtained through these simulations. This plot shows all dark matter models at $z = 8$, where one can observe some visual evidence of the damping of matter density fluctuations on small scales in the WDM scenarios. Also, note that the free streaming length $\lambda_{\rm fs}$ scales with the mass of the warm dark matter particle as $\lambda_{\rm fs} \propto \left(m_{\rm wdm}\right)^{-4/3} $. These scaling relations have been derived in \citep{2012MNRAS.424..684S, 2011PhRvD..84f3507S}. So, the lighter the WDM particle, the larger the free streaming scale, and the suppression will be more pronounced in 2 keV WDM model compared to 3 keV WDM model. This feature can also be seen in Figure~\ref{fig:odf} if we compare the matter overdensity and the halo mass field in the two WDM models.  
This suppression of density fluctuations at small scales will reduce the number of low mass haloes in the WDM scenarios. In Figure~\ref{fig:hmf}, we show the halo mass function obtained at three redshifts $z$ = $10$, $8$, and $6$ for all dark matter models. We observe that at any redshift, the number of low mass dark matter haloes gets reduced in the WDM model if we compare it to the $\Lambda$CDM model. Also, we observe that the suppression of low mass haloes is more for 2 keV WDM model compared to the 3 keV WDM model because the lighter the WDM particle, the larger the free streaming scale, and the formation of a more massive halo will be suppressed [see equation~(\ref{eq:M_FS})]. However, the number of massive haloes in the WDM model is very similar to the CDM model because, above the free streaming scale, the structure formation proceeds similarly to the CDM model. We have also shown the theoretically predicted Sheth-Tormen halo mass function estimated using a top-hat filter for all the dark matter models, where the transfer function in WDM scenarios is computed following the approach of \citet{2005PhRvD..71f3534V, 2013MNRAS.433.1573S}. 

\section{Simulating the redshifted 21-cm signal from the EoR} 
\label{method}
We have used the ReionYuga\footnote{\url{https://github.com/rajeshmondal18/ReionYuga}} semi-numerical simulations to simulate the redshifted 21-cm signal \citep{majumdar12,10.1093/mnras/stu1342, 2017MNRAS.464.2992M}. This simulation method is somewhat similar to the methods followed by \citet{2004ApJ...613....1F, 2009MNRAS.394..960C,zahn11, mesinger11}. Steps that are involved in this method can be summarized as follows: (I) Generating the dark matter density field, (II) Identifying the position and mass of the collapsed structures, i.e. haloes, in the dark matter field, (III) Assigning the ionizing photon production rates to these haloes, (IV) Generating the ionization maps including the effect of redshift space distortions and (V) Converting these ionization maps into redshifted 21-cm brightness temperature field.

As discussed in Section~\ref{stru}, we have used GADGET 2.0 N-body simulations to accomplish steps I and II. For the purpose of this work, we next assumed that the hydrogen follows the simulated dark matter distribution  at these redshifts (Table~\ref{tab:zx}). Further, as our reionization source model (step III), we assume that the number of ionizing photons produced by a halo is proportional to its mass \citep{2009MNRAS.394..960C, 10.1093/mnras/stu1342}:
\begin{equation}
    \label{eq:number_photons}
    N_\gamma (M) =  N_{\rm ion} \frac{M}{m_{\rm H}}\,,
\end{equation}
where $m_{\rm H}$ is the mass of a hydrogen atom, $M$ is the mass of the halo, and $N_{\rm ion}$ is a dimensionless free parameter, which may depend on various other degenerate reionization parameters including star-formation efficiency of a source, escape fraction of ionizing photons from these sources, etc. We set $N_{\rm ion} = 23.21$ to achieve $\bar{x}_{\HI} \approx 0.5$ at $z = 8$ in the $\Lambda$CDM model. This additionally ensures that for the $\Lambda$CDM scenario, reionization ends by $z \sim 6$ and produces a Thomson scattering optical depth that is consistent with the CMBR observations. 
\begin{table}
\centering
\caption{This tabulates the redshift $z$ and corresponding mass averaged neutral fraction $\bar{x}_{\HI}$ for all the dark matter models, for which we have simulated the 21-cm signal}
\label{tab:zx}
\begin{tabular}{cccc}
\hline
Redshift & $\bar{x}_{\HI}$ & $\bar{x}_{\HI}$ & $\bar{x}_{\HI}$\\
$z$ & (CDM) & (3 keV WDM) &  (2 keV WDM)\\ [0.5ex]
\hline
\hline
10.00 & 0.84 & 0.90 & 0.94\\
9.25 & 0.75 & 0.84 & 0.90\\
9.00 & 0.71 & 0.81 & 0.88\\
8.65 & 0.64 & 0.77 & 0.84\\
8.00 & 0.47 & 0.64 & 0.75\\
7.50 & 0.30 & 0.51 & 0.64\\
7.00 & 0.11 & 0.34 & 0.51\\
6.00 & 0.00 & 0.03 & 0.13\\
\hline
\end{tabular}
\end{table}

Next, we perform the step IV, i.e. we produce the ionization map, using the \HI\ density map and the ionizing photon density maps. We use an excursion-set based algorithm to produce the ionization map at the desired redshifts. In this formalism, we first map the hydrogen density and ionizing photon density fields on a coarser $128^3$ grid with a grid spacing $[0.07 \times 8] = 0.56$ Mpc. Next, to identify the ionized regions, we smooth both the \HI\ and ionizing photon fields using spheres of radius $R$ for $R_{\rm min} \leq R \leq R_{\rm max}$, where $R_{\rm min}$ is the resolution of the simulation which is equal to the grid spacing and $R_{\rm max}$ is the mean free path of the ionizing photon ($R_{\rm mfp}$). In these simulations, we keep the value of the mean free path of the ionizing photons $R_{\rm mfp} = 20$ Mpc, which follows from the findings of \cite{Songaila_2010}. For any grid point $\mathbf{x}$, if the averaged ionizing photon density $\langle n_{\gamma} (\mathbf{x}) \rangle_{\rm R}$ exceeds the averaged \HI\ density $\langle n_{\rm H} (\mathbf{x}) \rangle_{\rm R}$ for any $R$, then we flag that grid point to be ionized. The points which do not meet this criteria, we assign an ionization fraction $x_{\ion{H}{II}} (\mathbf{x})$ = $\langle n_{\gamma} (\mathbf {x}) \rangle_{\rm R_{\rm min}} / \langle n_{\rm H} (\mathbf{x}) \rangle_{\rm R_{\rm min}}$ to them. This is repeated for all grid points in the simulation volume and for all $R$ within the allowed range, and the ionization map is produced. This ionization is then converted into the 21-cm $\HI$ brightness temperature map (step V) following the equation~(\ref{eq:deltaTb}).
\begin{equation}
\label{eq:deltaTb}
\delta T_{\rm b} = 27 x_{\HI} \left(1 + \delta_{\rm b}\right) \left( \frac{\Omega_{\rm b} h^2}{0.023} \right) \left( \frac{0.15}{\Omega_{\rm m} h^2} \frac{1+z}{10} \right)^{1/2} \left( \frac{T_{\rm s} - T_{\gamma}}{T_{\rm s}} \right)\,, 
\end{equation}
where $x_{\HI}$ is the neutral hydrogen fraction, $\delta_{\rm b}$ is the fractional baryon overdensity, $T_{\rm s}$ and $T_\gamma$ are the spin temperature and the CMB temperature respectively. In this work, we assume that during reionization the IGM is substantially heated above the CMB ($T_{\rm s}\approx T_{\rm K}\gg T_{\gamma}$). This is a reasonable assumption once the global neutral fraction is $x_{\HI} \leq 0.9$ and has been independently observed in different studies of reionization \citep{2004ApJ...613....1F, 2009MNRAS.394..960C,majumdar12, 10.1093/mnras/stu1342, 2017MNRAS.464.2992M}. So, the term $(T_{\rm s} - T_{\gamma})/ T_{\rm s} \rightarrow 1$ in equation~(\ref{eq:deltaTb}). The matter peculiar velocities will make the redshifted 21-cm signal anisotropic along the line-of-sight of a present day observer. This unavoidable anisotropy in any cosmic signal is popularly known as the redshift space distortions. We follow the formalism of \citet{majumdar12} to map the real space brightness temperature field into the redshift space. 
\begin{figure}
    \centering
    \includegraphics[width=\columnwidth]{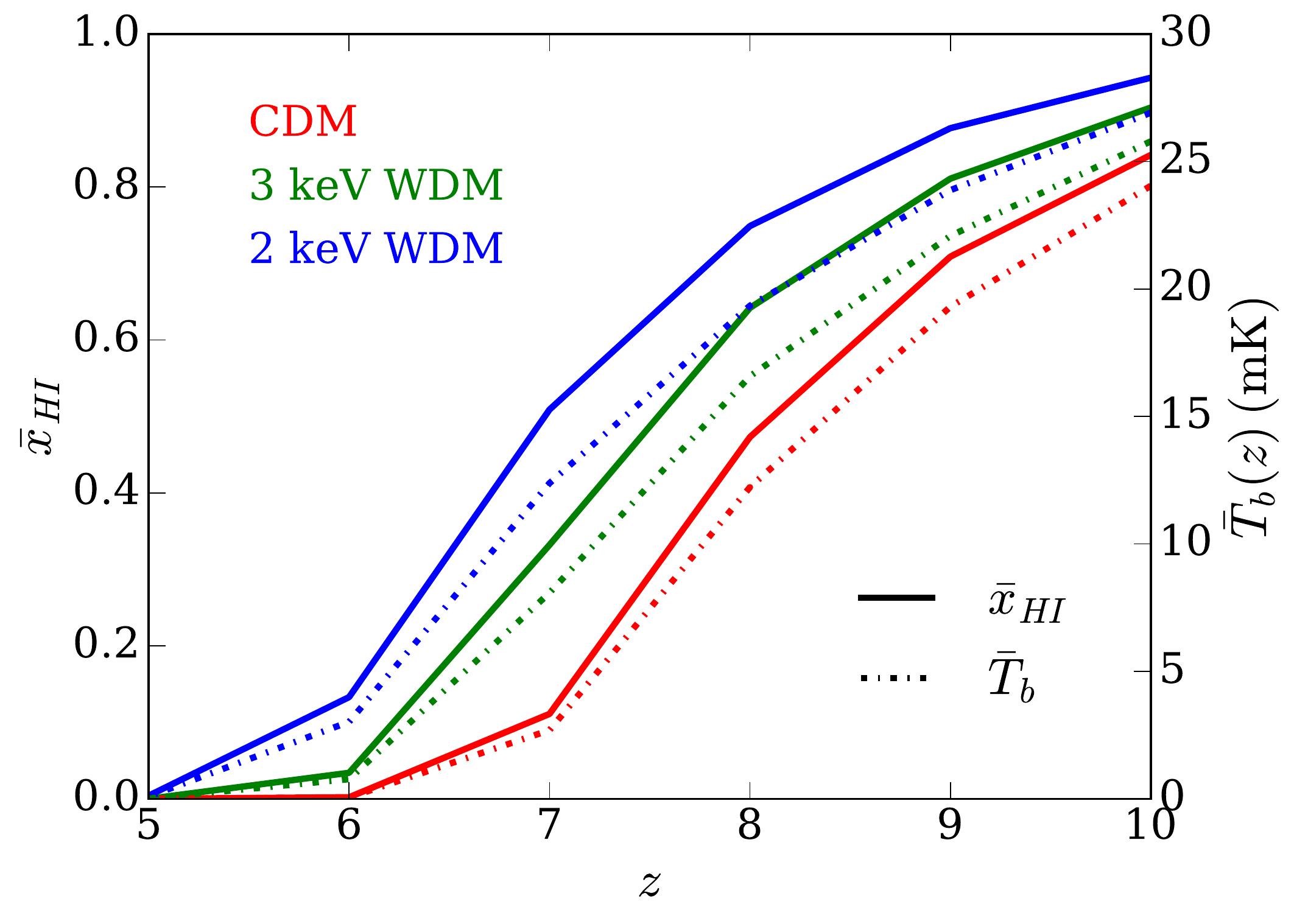}
    \caption{Reionization history for CDM (red), 3 keV WDM (green) and 2 keV WDM (blue) with the variation of mass averaged neutral fraction $\bar{x}_{\HI}$ (solid) and mean 21-cm brightness temperature $\bar{T}_{\rm b}$ (dotted-dashed) with redshift $z$.}
    \label{fig:history}
\end{figure}

Our simulation method is different from the similar approaches of \citet{2004ApJ...613....1F,zahn11, mesinger11}. The main differences are the following: we use a dark matter density field generated by an N-body simulation, whereas many of these methods use the matter densities from Zeldovich approximations; we use actual haloes as the host for the sources of reionization, whereas many of these methods use a normalized Press-Schechter formalism to directly calculate the collapsed fraction for different smoothing scales; for a specific smoothing scale if the ionization condition discussed earlier is satisfied, we paint only the centre of the smoothing sphere as ionized, whereas these other approaches paint the entire sphere as ionized; we use actual matter peculiar velocities to implement the redshift space distortions in the resulting 21-cm maps, whereas all of these methods use an approximate approach to implement the redshift space distortions. For a more detailed discussion on the issue of differences between different semi-numerical simulations of reionization we refer the interested reader to \citet{10.1093/mnras/stu1342}.
\begin{figure}
    \centering
    \includegraphics[scale = 0.25]{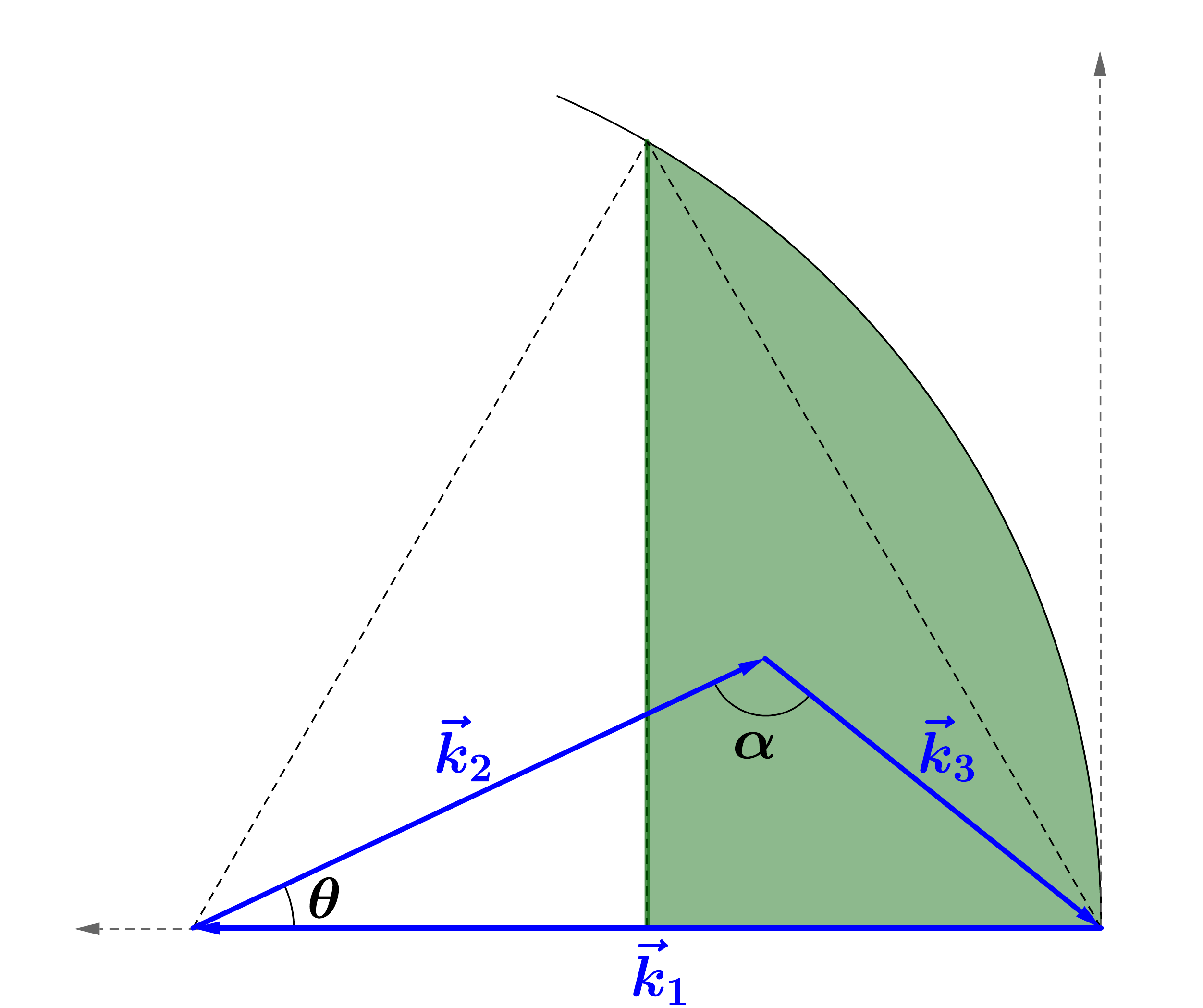}\\[0.5cm]
    \includegraphics[width=\columnwidth,angle=0]{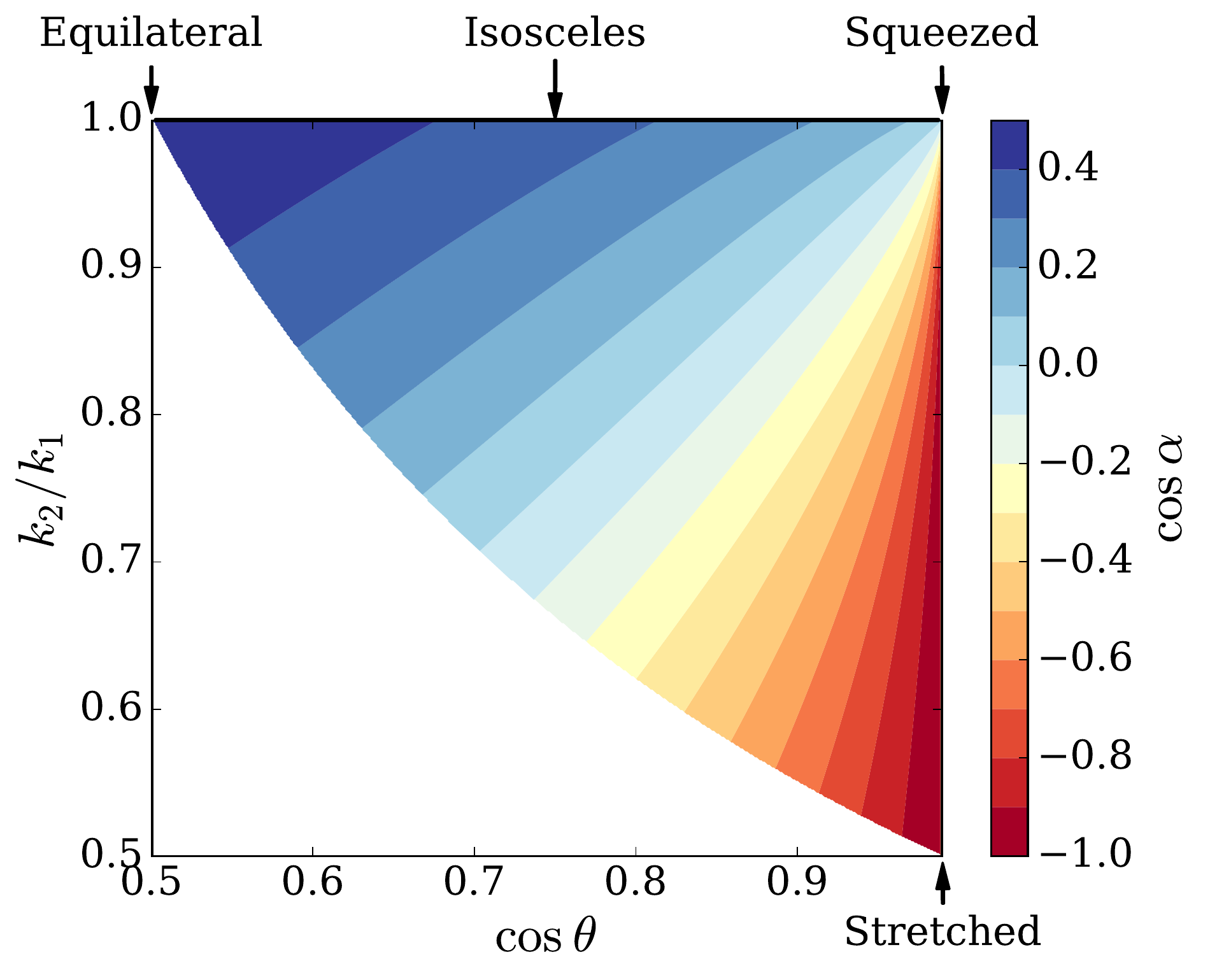}
    \caption{\textbf{Top panel:} For our unique triangle configurations, the tip of the $\mathbf{k_2}$ can move only in the green shaded region shown here, thereby satisfying equation~(\ref{eq:unique}). \textbf{Bottom panel:} Unique triangles parameter space in terms of $\cos{\alpha}$ showing the different limits of the $\mathbf{k}$-triangles \citep{2020MNRAS.tmp..281B}, for which we estimate the bispectra.}
    \label{fig:triangle}
\end{figure}
\begin{figure*}
    \centering
    \includegraphics[scale = 0.64]{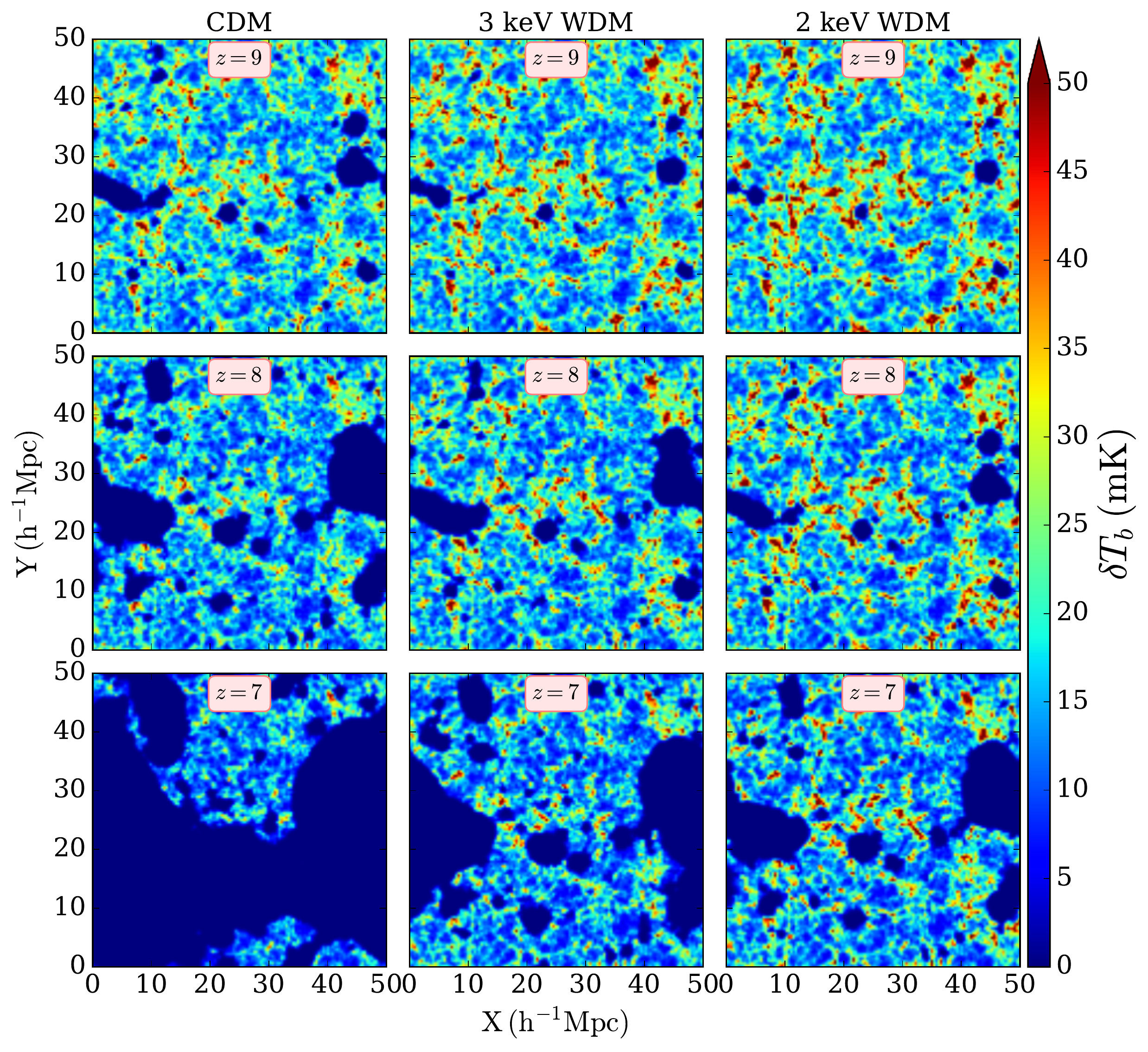}
    \caption{The three columns show the redshift space 21-cm $\HI$ brightness temperature maps for CDM, 3 keV WDM and 2 keV WDM respectively at $z$ = 9, 8 and 7 (from top to bottom). The color-bar represents the amplitude of 21-cm brightness temperature in mK.}
    \label{fig:HI_maps}
\end{figure*}

Figure~\ref{fig:history} shows the resulting reionization histories (i.e. $\bar{x}_{\HI} - z$ or $\bar{T}_b - z$ curves) produced from three different models of the dark matter. This figure clearly shows the expected delay in the reionization process (when one keeps the sources of ionization equally efficient in all cases) in the WDM scenarios compared to the CDM scenario. Figure~\ref{fig:HI_maps} shows the resulting \HI\ 21-cm maps for three different dark matter models at the same redshifts ($z = 9, 8$, and $7$). This also demonstrates the clear differences in \HI\, topology due to the differences in the structure formation history in these three cases.  

\subsection{Statistics of the simulated EoR 21-cm signal}
Visibilities, which are the Fourier transform of the sky brightness temperature, are the basic observables in any radio interferometric observation. This is one of the major reasons why most of the efforts to quantify the 21-cm signal are focused on using various Fourier statistics. For this work, we consider an idealistic scenario where there is no foreground emission or noise present in the data, and the visibilities contain only the Fourier transform of the signal brightness temperature $\Delta T_{\rm b} (\mathbf{k})$. The two Fourier statistics that are in our focus for this work are the power spectrum and bispectrum.
\subsubsection{\textit{21-cm power spectrum}}
One can define the 21-cm power spectrum as
\begin{equation}
\label{eq:Pk}
\langle \Delta T_{\rm b} (\mathbf{k})\, \Delta T_{\rm b}^{*} (\mathbf{k}^{'}) \rangle
 = (2 \upi)^3 \delta^3(\mathbf{k} - \mathbf{k}^{'})\, P(k)\,,
\end{equation}
where $\delta^3(\mathbf{k} - \mathbf{k}^{'})$ is the Dirac delta function, defined as
  \begin{equation}
    \delta^3 (\mathbf{k} - \mathbf{k}^{'}) =
    \begin{cases}
      1, & \text{if}\ \mathbf{k} = \mathbf{k}^{'} \\
      0, & \text{otherwise}\,.
    \end{cases}
  \end{equation}

To estimate the 21-cm power spectrum, we divide the entire $k$ range (determined by the smallest and largest length scales probed by our simulation) into $10$ equispaced logarithmic bins. Next, we Fourier transform the simulated 21-cm brightness temperature volume and use equation~(\ref{eq:Pk}) to estimate the power spectrum.

\subsubsection{\textit{21-cm bispectrum}}
Similar to the power spectrum the bispectrum can be defined as
\begin{equation}
\label{eq:Bk}
\langle \Delta T_{\rm b} (\mathbf{k_1}) \,\Delta T_{\rm b} (\mathbf{k_2}) \,\Delta T_{\rm b} (\mathbf{k_3}) \rangle = V \delta^3_{\mathbf{k_1} + \mathbf{k_2} + \mathbf{k_3},\,0} \,B_{\rm b}(\mathbf{k_1}, \,\mathbf{k_2}, \,\mathbf{k_3})\,,
\end{equation}
where  $\delta^3_{\mathbf{k_1} + \mathbf{k_2} + \mathbf{k_3},\,0}$ is the Dirac Delta function that ensures that the three ${\mathbf k}$ modes involved form a closed triangle (see top panel of Figure~\ref{fig:triangle}) i.e.
  \begin{equation}
    \delta^3_{\mathbf{k_1} + \mathbf{k_2} + \mathbf{k_3},\,0} =
    \begin{cases}
      1, & \text{if}\ (\mathbf{k_1} + \mathbf{k_2} + \mathbf{k_3}) = 0 \\
      0, & \text{otherwise}\,.
    \end{cases}
  \end{equation}
  
Following the definition of the bispectrum one can define a binned estimator for this statistic as
\begin{equation}
\label{eq:Bk_est}
\hat{B}(\mathbf{k_1}, \,\mathbf{k_2}, \,\mathbf{k_3}) = \frac{1}{N\,V} \mathlarger{\sum}_{(\mathbf{k_1} + \mathbf{k_2} + \mathbf{k_3} = 0) \,\in \,n}  \Delta T_{\rm b} (\mathbf{k_1}) \,\Delta T_{\rm b} (\mathbf{k_2}) \,\Delta T_{\rm b} (\mathbf{k_3})\,,     
\end{equation}
where the total number of triangles that belong to the $n^{th}$ triangle configuration bin is $N$, and $V$ is the simulation volume.  

To estimate the bispectra from the simulated signal we follow the algorithm of \citet{2018MNRAS.476.4007M}. In their algorithm \citet{2018MNRAS.476.4007M} have parametrized the triangle configurations using two independent parameters:\\
    (1) The ratio between the length of the two arms of a ${\mathbf k}$ triangle is
\begin{equation}
\label{eq:k2_k1}
    k_2/k_1 = n\,.
\end{equation}
    (2) The cosine of the angle between these two arms is given by
\begin{equation}
\label{eq:cos_theta}
    \cos \theta = \frac{\mathbf{k_1} \cdot \mathbf{k_2}}{k_1 k_2}\,.
\end{equation}
This parametrization helps in reducing the overall computation time for bispectrum estimation.

For a comprehensive understanding of the 21-cm bispectra, we need to estimate the bispectrum for all possible unique triangle configurations. We follow the prescription of \citet{2020MNRAS.tmp..281B}, that allows us to identify all possible unique triangles using the two parameters defined  in equations \eqref{eq:k2_k1} and \eqref{eq:cos_theta} and by imposing the additional condition on the triangle parameter space i.e.
\begin{equation}
\label{eq:unique}
    k_1 \geq k_2 \geq k_3 \implies \frac{k_2}{k_1}\cos{\theta} \geq 0.5\,.
\end{equation}
Figure~\ref{fig:triangle} shows the unique triangle configurations in the $n-\cos{\theta}$ parameter space. 

\section{Results} \label{results}
\subsection{Reionization history and 21-cm topology}
The differences in the dark matter models have two prominent impacts on the  21-cm signal from the EoR. One of them is the delay in the global reionization process, which is evident from the $\bar{x}_{\HI}-z$ and $\bar{T}_{\rm b} - z$ curves for different dark matter models in Figure~\ref{fig:history}. The reionization starts and finishes later in the WDM scenarios compared to the CDM scenario. This is a clear signature of the differences in the structure formation history in different scenarios (shown through the halo mass functions in Figure~\ref{fig:hmf}). The halo mass functions plotted in Figure~\ref{fig:hmf} reveals that the low mass haloes cumulatively are the dominant sources of ionizing photons (when we assume all haloes are equally efficient in producing ionizing photons) at any redshift for all dark matter models considered here. In the case of warm dark matter models, the low mass end of the halo mass function gets suppressed compared to the cold dark matter scenario, which results in an overall decrease in the total number of ionizing photons produced at any redshift. This overall reduction in the total number of ionizing photons produced at any redshift leads to a delay in the reionization process for the WDM models compared to the CDM one. The suppression of the low mass end of the halo mass function is more pronounced in the case of the $2$ keV WDM model compared to the $3$ keV WDM model, which leads to an even longer delay in the reionization process in case of the $2$ keV WDM model. 

The other prominent impact that the differences in the dark matter models on the reionization process have is the differences in the \HI\ 21-cm brightness temperature topology. This is also caused by the different levels of suppression of the low mass end of the halo mass function in different dark matter models. Figure~\ref{fig:HI_maps} shows two-dimensional slices of the \HI\ 21-cm brightness temperature maps at three different redshifts: $z$ = $9$, $8$ and $7$ for all dark matter models. In these figures, a clear difference in the size and location of the ionized regions is visible in different dark matter models. However, it is important to note that even if the 21-cm maps are shown at the same redshifts, they are not at the same state of reionization for different dark matter models. One may ask how much of this observed difference in the 21-cm topology is due to the delay in reionization history. To address this, \citet{2018JCAP...08..045D} have tuned the ionizing photon production efficiency (i.e. the parameter $N_{\rm ion}$ in our simulations) of haloes in different dark matter models to get the same global neutral fraction at the same redshifts for all dark matter models. With this modification in their simulations, they still observed significant differences in the 21-cm topology and its Fourier statistics for different dark matter scenarios. However, one should note that changing the $N_{\rm ion}$ in different scenarios effectively means changing the reionization model altogether. Therefore, for the main analysis and results in this paper, we keep the value of $N_{\rm ion}$ same for all dark matter scenarios. We briefly discuss the impact of changing the $N_{\rm ion}$ (to get the same reionization history) for different dark matter models on the 21-cm statistics (observables) in Appendix~\ref{changednion}.

\subsection{21-cm power spectrum}
\begin{figure}
    \centering
    \includegraphics[width = \columnwidth]{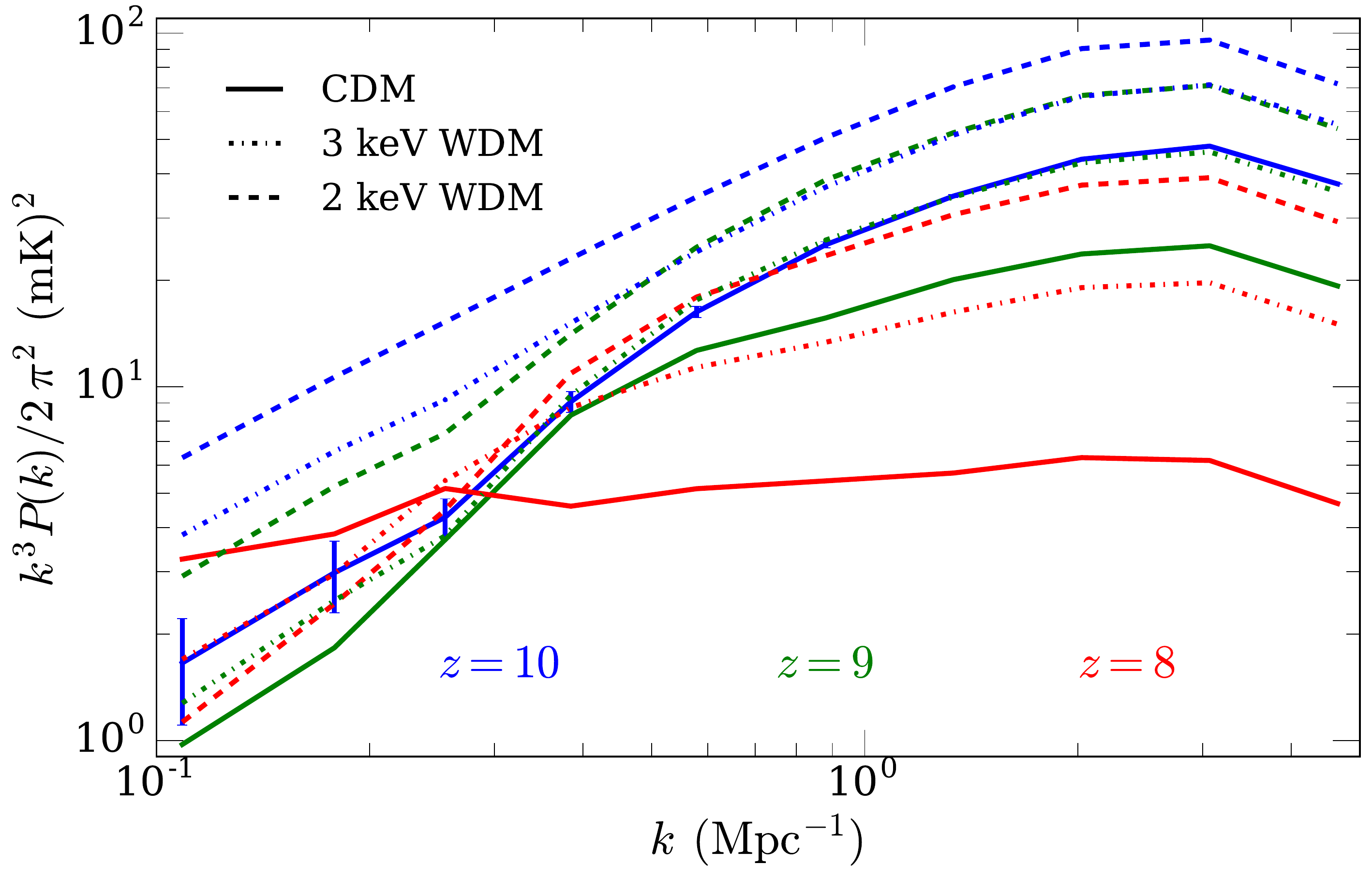}\\[0.10cm]
    \includegraphics[width = \columnwidth]{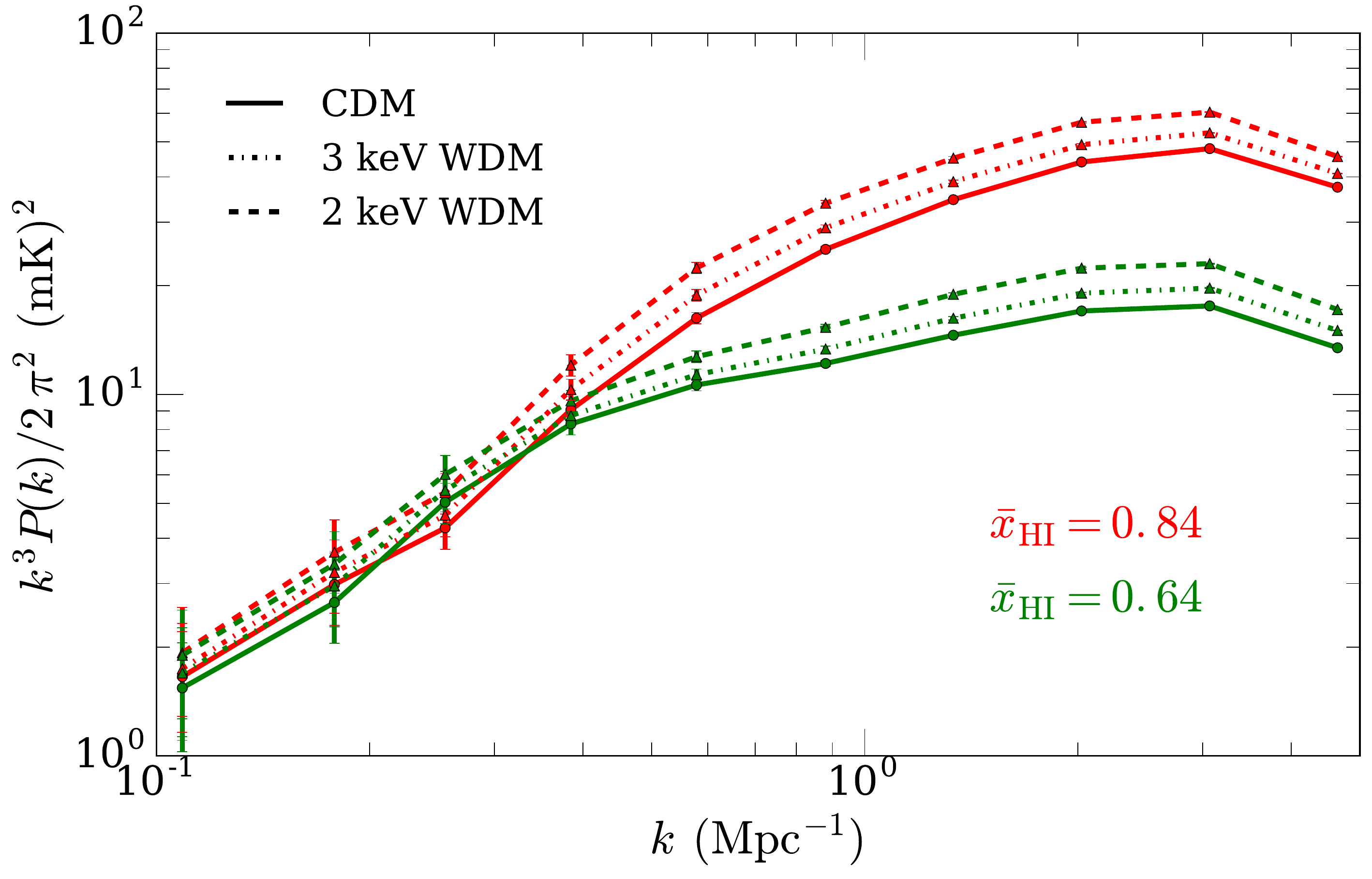}\\[0.10cm]
    \includegraphics[width = \columnwidth]{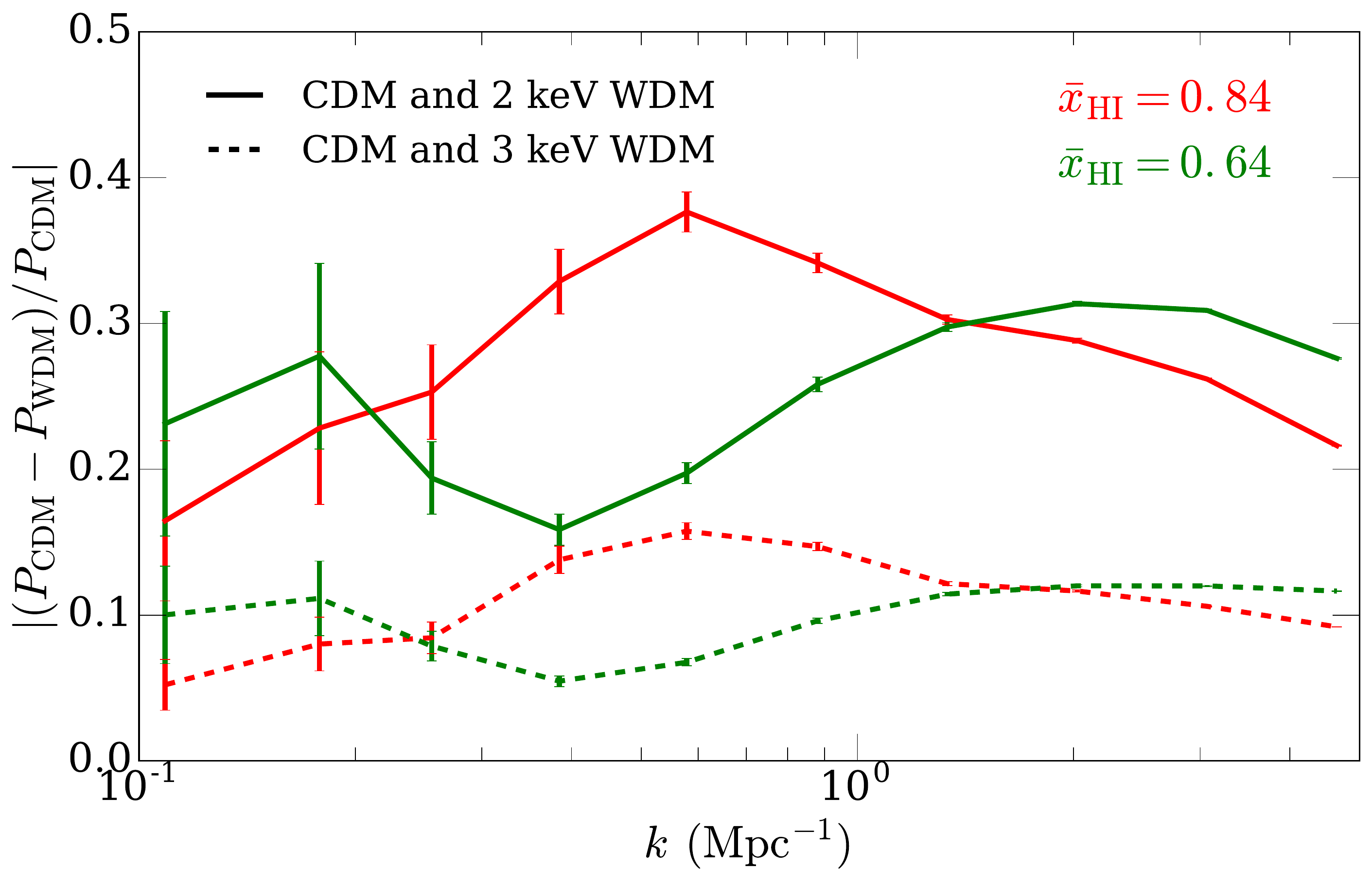}
    \caption{\textbf{Top panel:} Simulated 21-cm power spectrum for CDM (solid), 3 keV WDM (dotted-dashed) and 2 keV WDM (dashed) at $z$ = 10 (blue), 9 (green) and 8 (red). \textbf{Central Panel:} 21-cm power spectrum for CDM (solid), 3 keV WDM (dotted-dashed) and 2 keV WDM (dashed) at same mass averaged neutral fraction for $\bar{x}_{\HI}$ = 0.84 (red) and 0.64 (green). \textbf{Bottom Panel:} Relative fractional difference in 21-cm power spectrum between CDM and 2 keV WDM (solid), and between CDM and 3 keV WDM (dashed) estimated at $\bar{x}_{\HI}$ = 0.84 (red) and 0.64 (green).}
    \label{fig:Pk}
\end{figure}
We next focus our attention to the statistic that one would use to detect the EoR 21-cm signal using radio interferometers, i.e. the power spectrum. The power spectrum quantifies the amplitude of fluctuations in the signal at different length scales. The top panel of Figure~\ref{fig:Pk} shows the power spectrum in all three dark matter scenarios (represented using three different line styles) at three different redshifts (represented using three different line colours). The evolution of the power spectra with redshift for a specific dark matter model demonstrated in Figure~\ref{fig:Pk} follows the behaviour of an inside-out reionization \citep{2009MNRAS.394..960C, mesinger11, zahn11, 10.1093/mnras/stu1342, 2017MNRAS.464.2992M}, i.e.\ ionization starts at the highest density regions (where the ionizing sources are) in the IGM and then it gradually makes it's way to the low density regions. The kind of inside-out reionization makes power at the large scales (small $k$ modes) grow in amplitude as reionization progresses and reach its peak at $\bar{x}_{\HI}\sim 0.5$ and then go down in amplitude. The top panel of Figure~\ref{fig:Pk} clearly shows that the difference in the 21-cm power spectra for different dark matter models is quite large (by even few orders of magnitude in certain $k$ modes), when they are compared at the same redshifts. However, even if these power spectra are at the same redshifts, they are from different stages of reionization for different dark matter models. To get a better idea of how a dark matter model impacts the reionization process it is more reasonable to compare the 21-cm statistics at the same stage of reionization (determined by the $\bar{x}_{\HI}$ value) in different dark matter models.

\subsubsection{\textit{Difference in 21-cm power spectrum between different models at same mass averaged neutral fraction}}
The amplitude of the redshifted 21-cm power spectrum during the EoR is determined mainly by the fluctuations in the neutral fraction ($\bar{x}_{\HI}\,$) field. This is why it is more logical to compare the EoR 21-cm power spectra from different dark matter models at the same stages of reionization (i.e. approximately for the same values of $\bar{x}_{\HI}\,$). Here the expectation is when compared for the same values of $\bar{x}_{\HI}\,$, the differences in $P(k)$ will be mainly due to differences in the 21-cm topology, rather than the overall level of ionization of the IGM or the delay in the reionization history. The central panel of Figure~\ref{fig:Pk} shows the 21-cm power spectrum at approximately same averaged neutral fraction values ($\bar{x}_{\HI} = \{0.84,\, 0.64\}$) for different dark matter models (the corresponding redshifts for different dark matter models are tabulated in Table~\ref{tab:zx}). It is obvious from this plot that the differences in power spectra between the WDM models and the CDM model become significantly low when compared in this way, and for small $k$ modes (or large length scales), these differences are within the sample variance limits. 

To quantify the differences between the EoR 21-cm power spectra from WDM and CDM scenarios, we estimate the quantity $|(P_{{\rm CDM}}-P_{{\rm WDM}})/P_{{\rm CDM}}|$. The bottom panel of Figure~\ref{fig:Pk} shows this relative fractional difference in 21-cm power spectrum estimated at same neutral fraction $\bar{x}_{\HI} = 0.84$ and $0.64$ between the WDM models and the CDM model. It is clear from this figure that the amplitude of this relative difference is larger for the $2$ keV WDM model compared to the $3$ keV WDM model.  During the early stage of reionization when $\bar{x}_{\HI} = 0.84$, this difference peaks around intermediate length scales and during the intermediate stage of reionization when $\bar{x}_{\HI} = 0.64$ it is peaked around small length scales. We observe that for a large range of $k$-modes, the relative fractional difference varies in the range $0.15 - 0.35$ for the comparison between the CDM and the $2$ keV WDM, and in the range $0.05 - 0.15$ for the comparison between the CDM and the $3$ keV WDM model. These differences are probably not large enough to be detectable by the first generation of radio interferometers, which have lower sensitivity. These results are consistent with the previous studies made by \citet{2018JCAP...08..045D}. Thus, the power spectrum is not an ideal tool to probe the differences between different dark matter models.

\subsection{21-cm bispectrum}
The power spectrum is an incomplete statistics when it comes to optimally quantifying the EoR 21-cm signal fluctuations, as this signal is highly non-Gaussian. We use the next higher-order statistics, the bispectrum, to quantify this evolving non-Gaussianity in the EoR 21-cm signal. The source of the non-Gaussianity in the signal is the non-random distribution of growing ionized regions in the IGM, which drives the fluctuations in the signal and also leads to the signal correlation between different length scales.

We first show the signal bispectrum for two specific triangle configurations, namely, equilateral and isosceles triangles.
\begin{figure}
    \centering
    \includegraphics[width = \columnwidth]{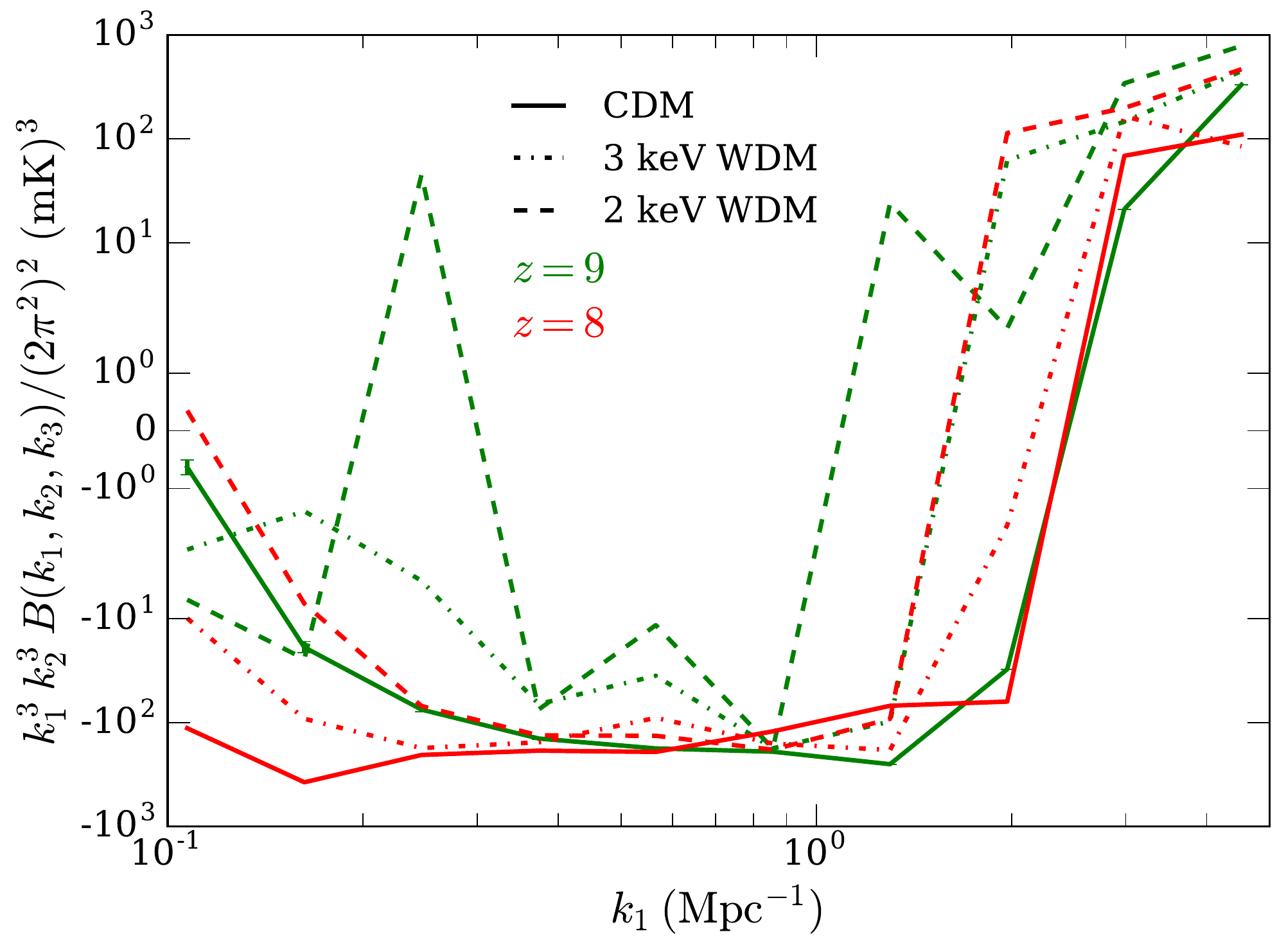}
    \includegraphics[width = \columnwidth]{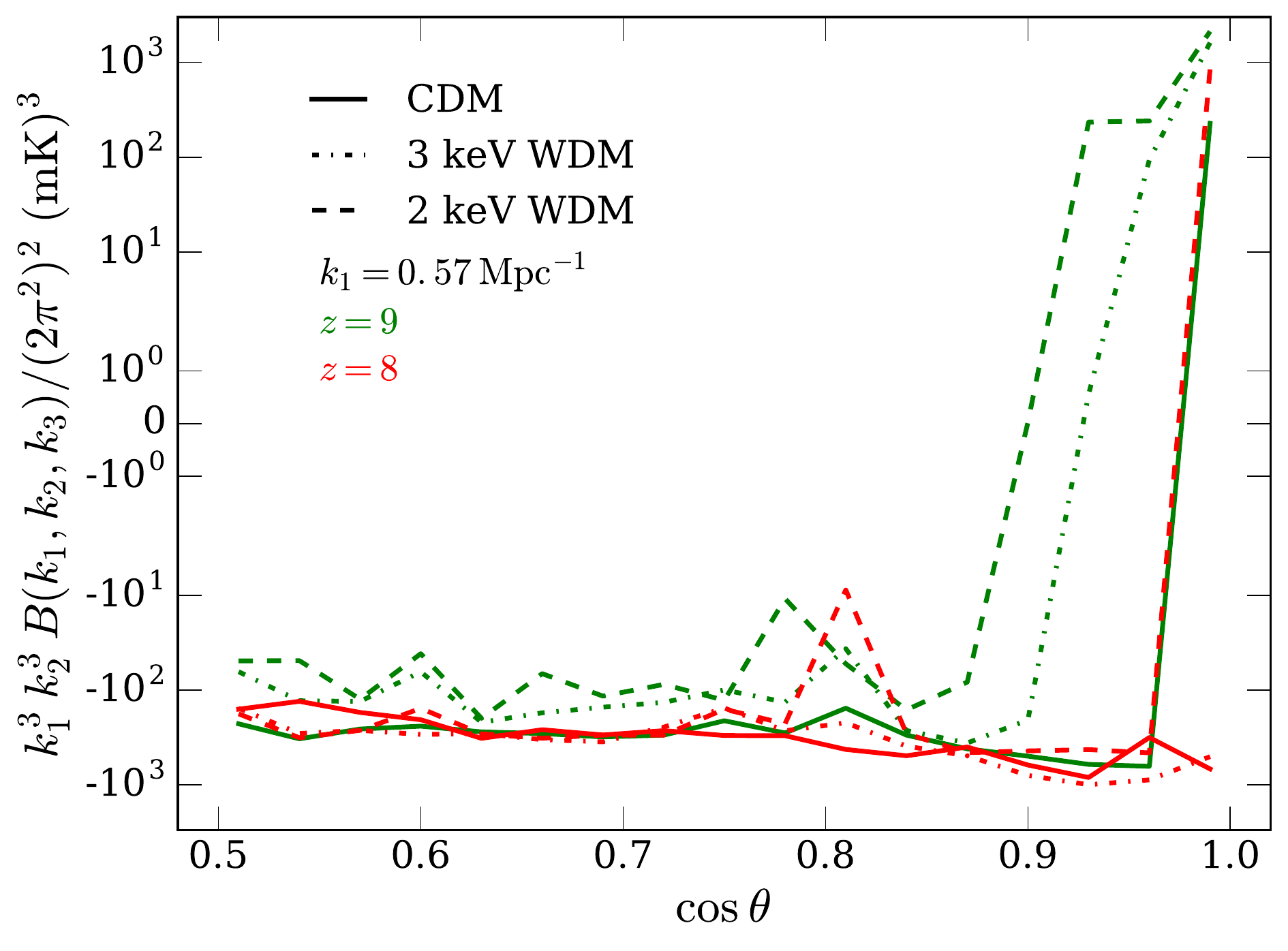}
    \caption{\textbf{Top panel:} Simulated 21-cm bispectrum for equilateral triangles for CDM (solid), 3 keV WDM (dotted-dashed) and 2 keV WDM (dashed) model at $z$ = 9 (green) and 8 (red). \textbf{Bottom Panel:} 21-cm bispectrum for isosceles triangles at $k_1 = 0.57$ Mpc$^{-1}$ for CDM (solid), 3 keV WDM (dotted-dashed) and 2 keV WDM (dashed) model at $z$ = 9 (green) and 8 (red).}
    \label{fig:bk_equil}
\end{figure}
The top panel of Figure~\ref{fig:bk_equil} shows the signal bispectrum for equilateral triangles (i.e.\ $k_1 = k_2 = k_3$) at $z$ = 9 and 8 for all three dark matter models. One can notice that for all the models, at smaller $k_1$ modes, the bispectra remains mostly negative and as we go towards larger $k_1$ modes, it changes its sign and becomes positive. In the bottom panel of Figure~\ref{fig:bk_equil}, we show the bispectra for isosceles triangles (i.e.\ $k_1 = k_2$ and $0.5 \leq \cos{\theta} < 1.0$) at $k_1 = 0.57$ Mpc$^{-1}$ for the same redshifts as the top panel. It is clear from this figure that the bispectra is negative for most of the $\cos{\theta}$ range and it changes sign and becomes positive around $\cos{\theta} \sim 0.9 $ for all dark matter models. For all of the dark matter models it reaches its maximum amplitude at the squeezed limit i.e. $\cos{\theta} \sim 1$. For both types of triangles one can also visually recognize the differences in the bispectra for different dark matter models. However, quantifying these differences from this plot is difficult as the range on the y-axis is very large.
\begin{figure}
    \centering
    \includegraphics[width = 0.965\columnwidth]{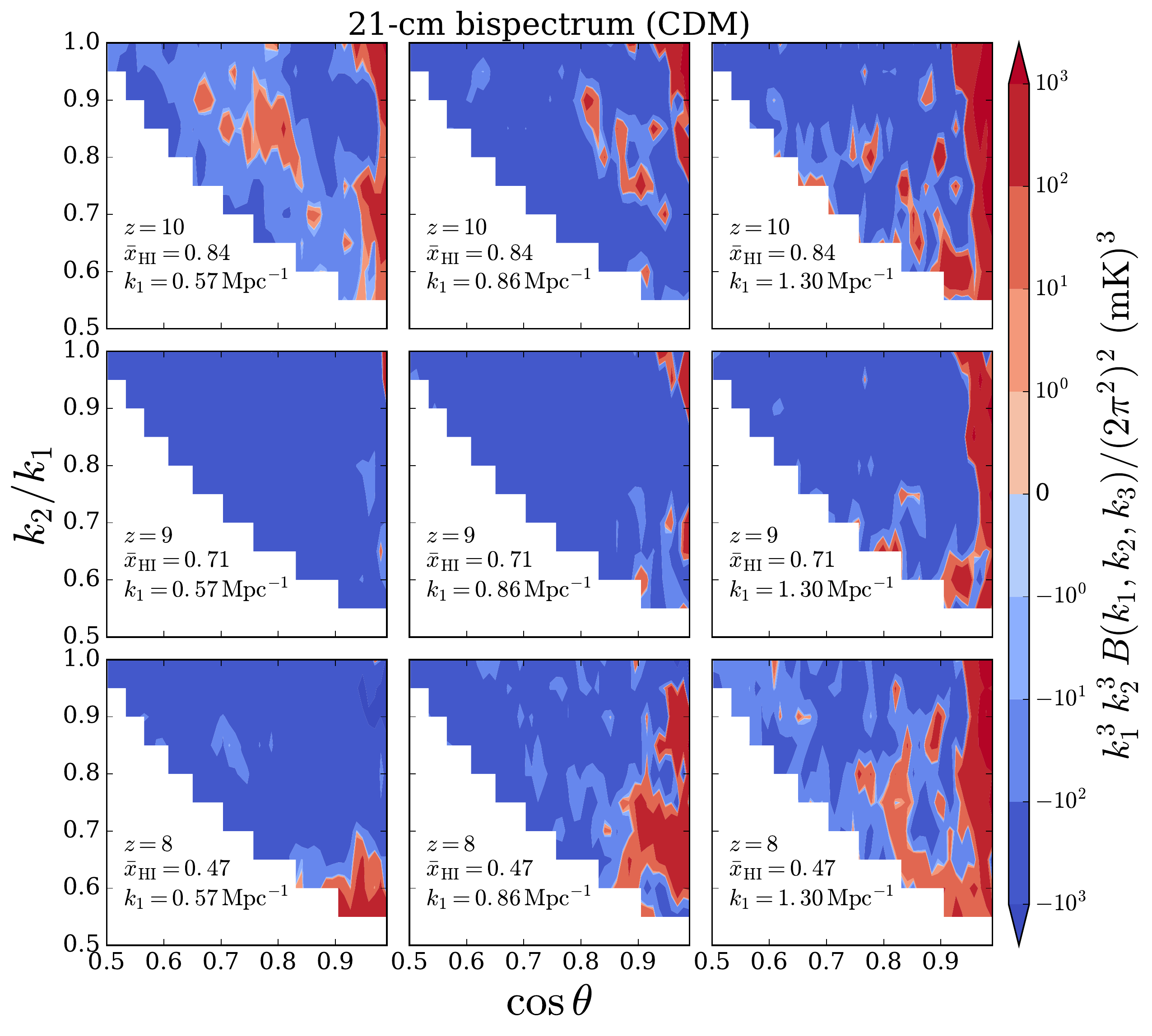}
    \includegraphics[width = 0.965\columnwidth]{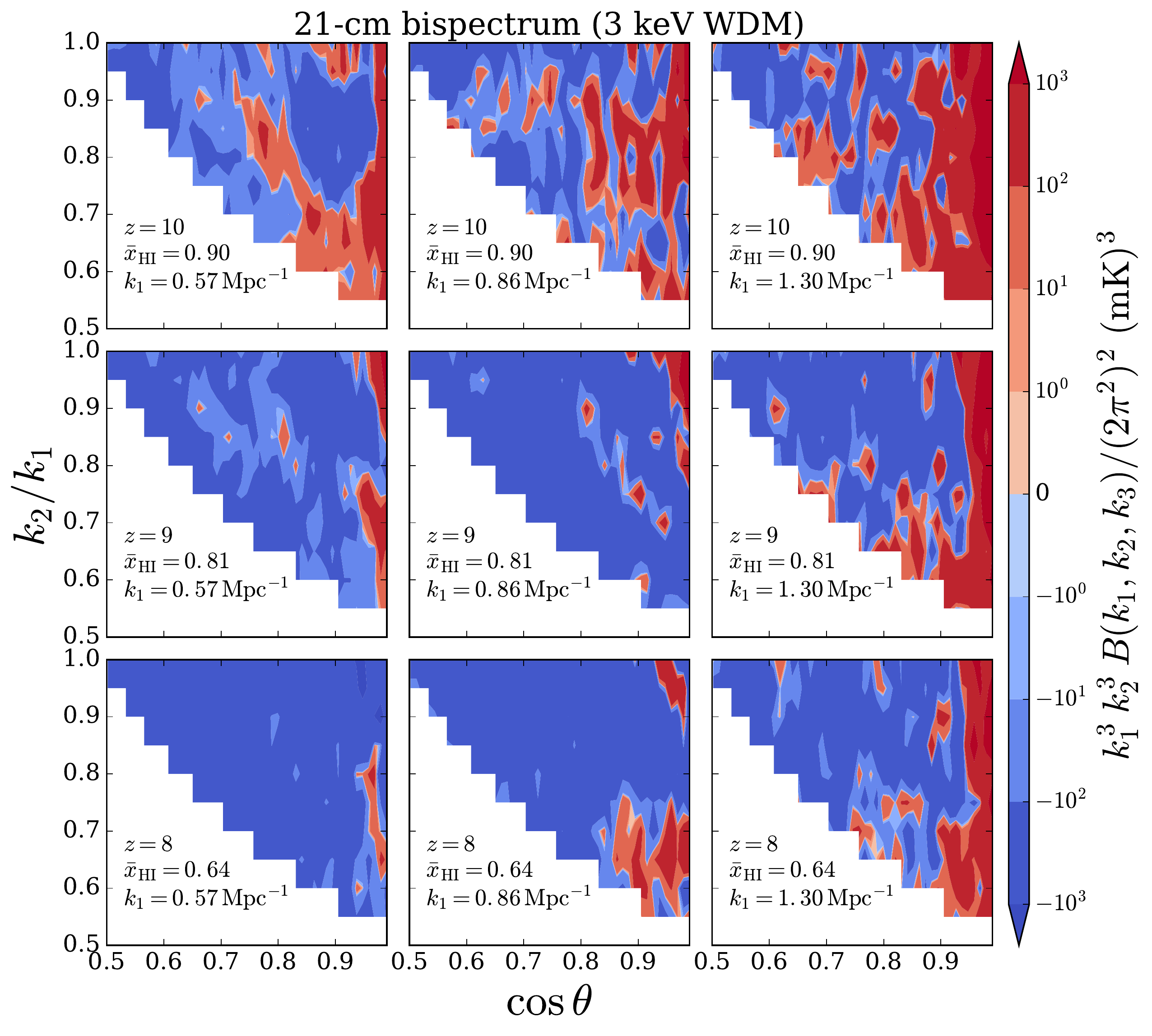}
    \includegraphics[width = 0.965\columnwidth]{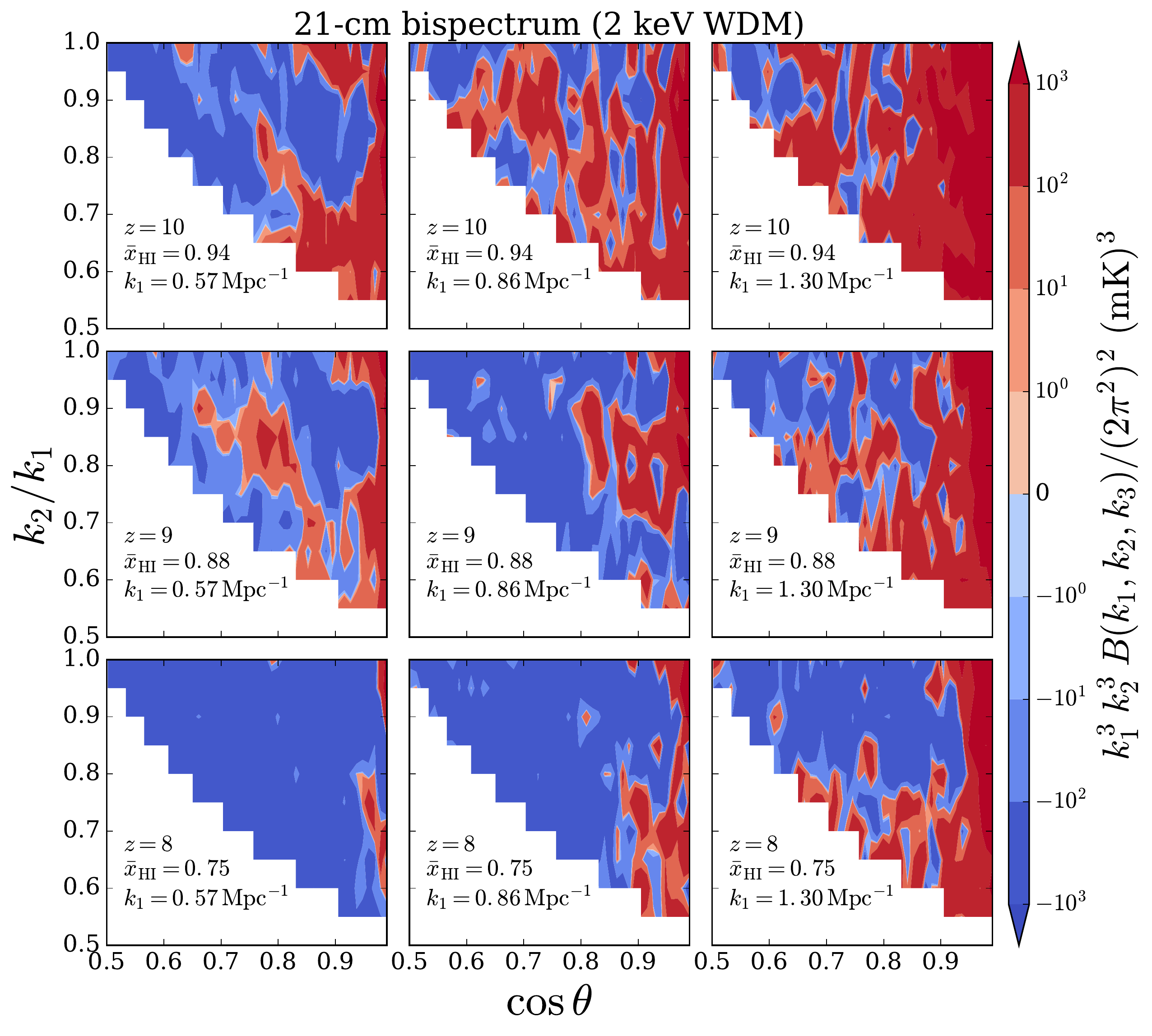}
    \caption{21-cm bispectrum for all unique triangles at $z$ = 10, 9 and 8 (top to bottom), and for $k_{1}$ = 0.57, 0.86 and 1.30 Mpc$^{-1}$ (left to right). Top panel: CDM; central panel: 3 keV WDM; bottom panel: 2 keV WDM model. The neutral fraction at these $z$ is shown in the figure.}
    \label{fig:Bk}
\end{figure}

Next, we show the signal bispectrum for all unique $k$-triangles (see Figure~\ref{fig:triangle}) for dark matter models to quantify this non-Gaussianity. Figure~\ref{fig:Bk} shows the bispectra at three redshifts $z$ = $\{10,~9,~8\}$, and for triangles with three $k_1$ modes $k_{1} = \{0.57,~0.86,~1.30\}$ Mpc$^{-1}$. We choose not to show the bispectra for $k_{1} < 0.57$ Mpc$^{-1}$ because of the high sample variance in these triangle bins (due to our small simulation volume). In an earlier study of the EoR 21-cm bispectrum (in real space) \citet{2018MNRAS.476.4007M} has shown that the bispectrum is mostly negative for triangles involving small $k$ modes. It has a maximum amplitude for the squeezed limit triangles \citep{2018MNRAS.476.4007M,2020MNRAS.492..653H}. The bispectrum also shows a very interesting feature; it changes its sign when one gradually moves from smaller $k$-triangles to triangles with larger $k$ modes. This makes the EoR 21-cm bispectra an even more interesting statistic for a confirmative detection of the signal \citep{2018MNRAS.476.4007M,2020MNRAS.492..653H}. However, all of these analyses were based on a few specific types of $k$-triangles and for the signal in real space. For a thorough analysis of the EoR 21-cm bispectra for all unique $k$-triangles in redshift space, the readers are requested to refer to Majumdar et al. (in prep.).

Based on the plots in Figure~\ref{fig:Bk} and the analysis of \citet{2018MNRAS.476.4007M} and Majumdar et al. (in prep) one can identify few more generic features of the inside-out EoR 21-cm bispectra in the entire unique $k$-triangle space (defined by parameters $n$ and $\cos{\theta}$), irrespective of the underlying dark matter model. At the very beginning of reionization ($\bar{x}_{\HI} \geq 0.90$) for smaller $k_1$-triangles bispectra for a significant fraction of the $n - \cos{\theta}$ parameter space is positive and in the rest of the parameter space it is negative. For the small $k_1$-triangles as reionization progresses bispectra in most of the $n - \cos{\theta}$ parameter space becomes negative ($0.85 \geq \bar{x}_{\HI} \geq 0.60$) and stays negative until the $\bar{x}_{\HI} \leq 0.40 $. At neutral fractions lower than this, the bispectrum starts to become positive again in a significant portion of the triangle parameter space. Further, for a fixed $\bar{x}_{\HI}$ value as one gradually moves from smaller to larger $k_1$-triangles, an even larger fraction of the $n - \cos{\theta}$ space starts to have positive bispectrum. The bispectra start to become positive, mainly near the squeezed and linear limits ($\cos{\theta} \simeq 1$) of triangles at almost during all stages of the reionization. The amplitude of the bispectra for small $k_1$-triangles reaches a maximum when $\bar{x}_{\HI} \sim 0.5$. The evolution of sign and amplitude of the bispectra can be interpreted in the light of the quasi-linear model of brightness temperature fluctuations \citep{mao12}. Using this model, one can show that there are total eight component bispectra (two auto and six cross) that contribute to the redshift space EoR 21-cm bispectra. Among these eight components, the most important ones for the small $k_1$-triangles are the two auto bispectrum of the \HI\ and the matter density field and three cross-bispectra between \HI\ and matter density fields. These components make the 21-cm bispectra negative in most of the parameter space. The contribution from the auto bispectrum of the matter density field is negligible for small $k_1$-triangles. As one moves towards the bispectra for large $k_1$-triangles contribution from the auto bispectrum of the matter density field (which is always positive in the sign for all types of triangles) grows and becomes significantly large at largest $k_1$-triangles and make the 21-cm bispectra positive. For a more detailed discussion on this topic, we refer the reader to Majumdar et al. (in prep).    

Figure~\ref{fig:Bk} clearly shows that if one compares the bispectra for different dark matter models at the same redshifts, as expected, the differences between them would be quite large. For some $k$-triangles and depending on the stages of the reionization, these relative differences can be even larger than the ones observed in the case of power spectra. This is because as the stages of the reionization are different for different models, the overall level of fluctuations, as well as the topology, will be significantly different, both of which affect the bispectrum amplitude and sign.

\subsubsection{\textit{Difference in 21-cm bispectrum between different models at same mass averaged neutral fraction}}
\begin{figure}
    \centering
    \includegraphics[width = \columnwidth]{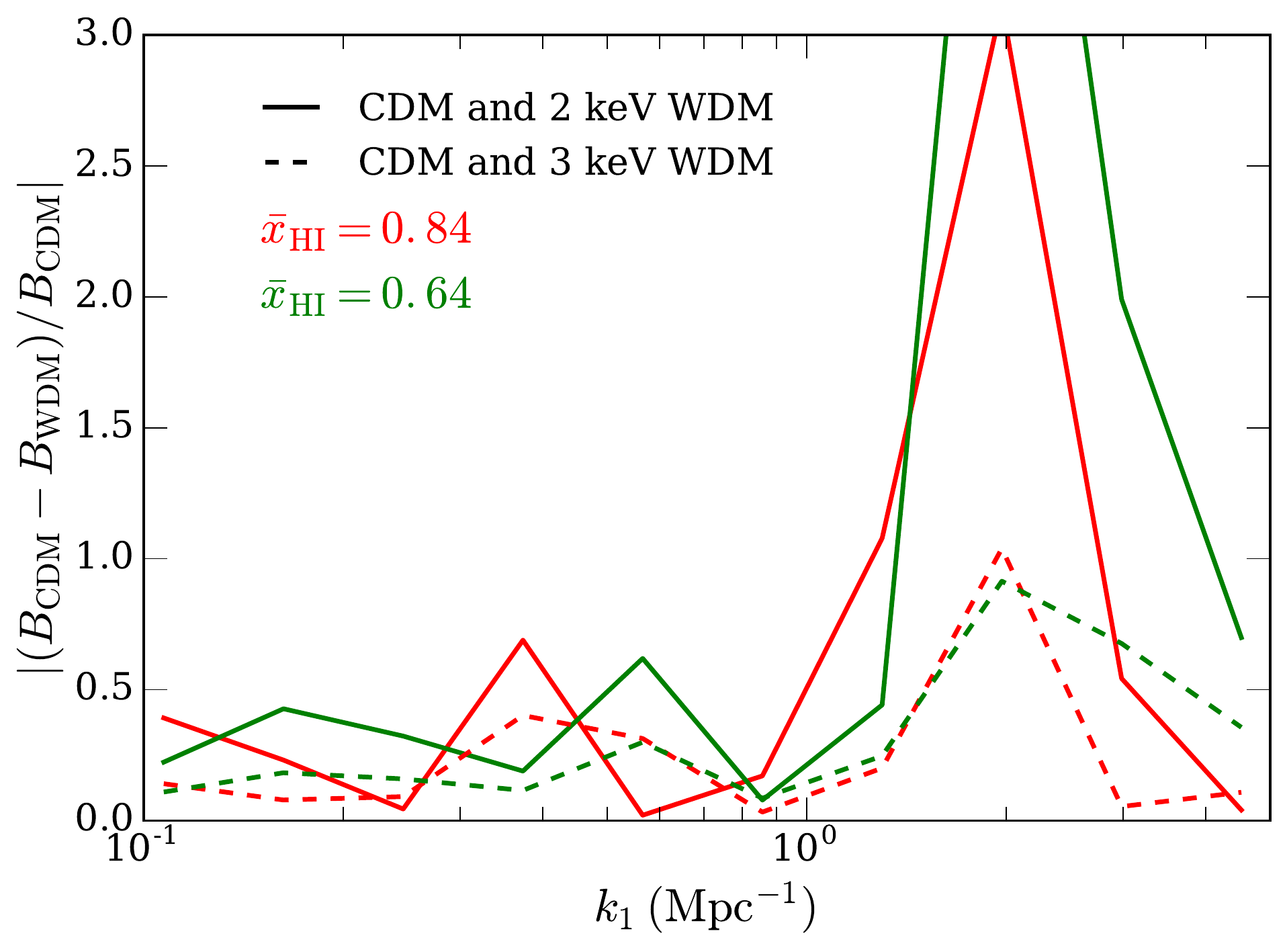}
    \includegraphics[width = \columnwidth]{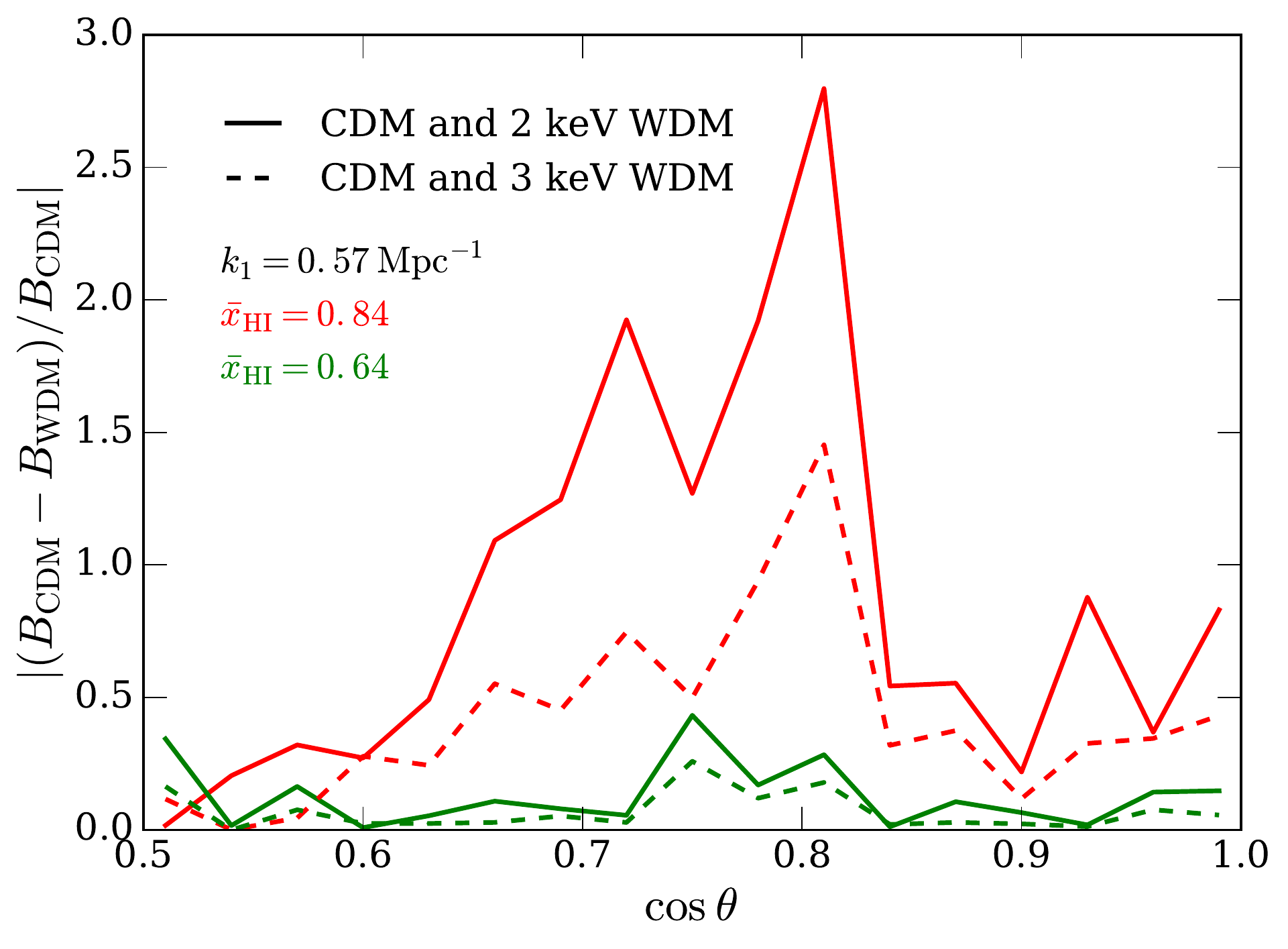}
    \caption{\textbf{Top panel:} Relative fraction difference in 21-cm bispectrum for equilateral triangles between CDM and 2 keV WDM model (solid), and between CDM and 3 keV WDM model (dashed) at $\bar{x}_{\HI}$ = 0.84 (red) and 0.64 (green). \textbf{Bottom Panel:} Relative fraction difference in 21-cm bispectrum for isosceles triangles at $k_1 = 0.57$ Mpc$^{-1}$ between CDM and 2 keV WDM model (solid), and between CDM and 3 keV WDM model (dashed) at $\bar{x}_{\HI}$ = 0.84 (red) and 0.64 (green).}
    \label{fig:bk_isosceles}
\end{figure}
\begin{figure}
    \centering
    \includegraphics[width = \columnwidth]{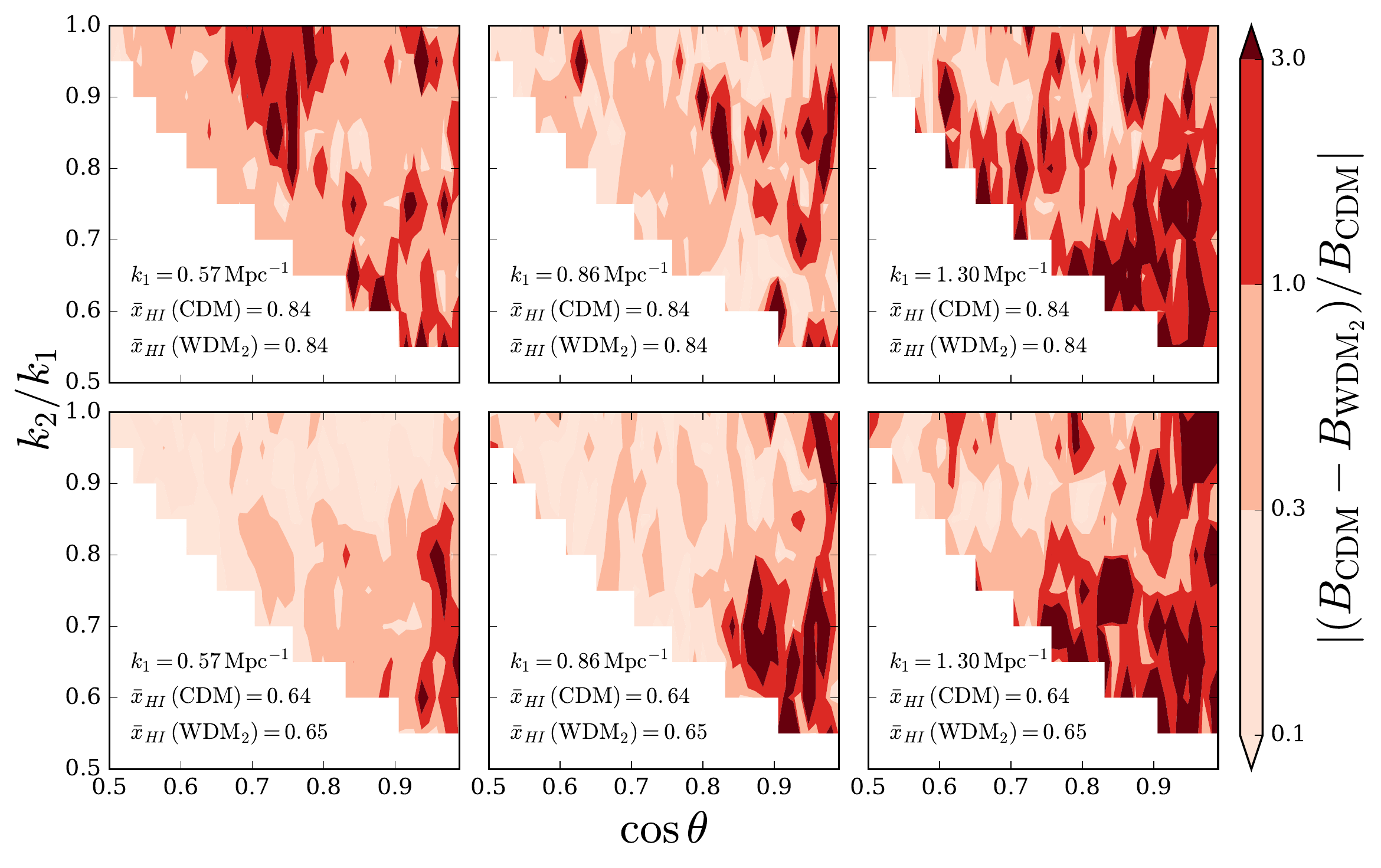}\\[0.3cm]
    \includegraphics[width = \columnwidth]{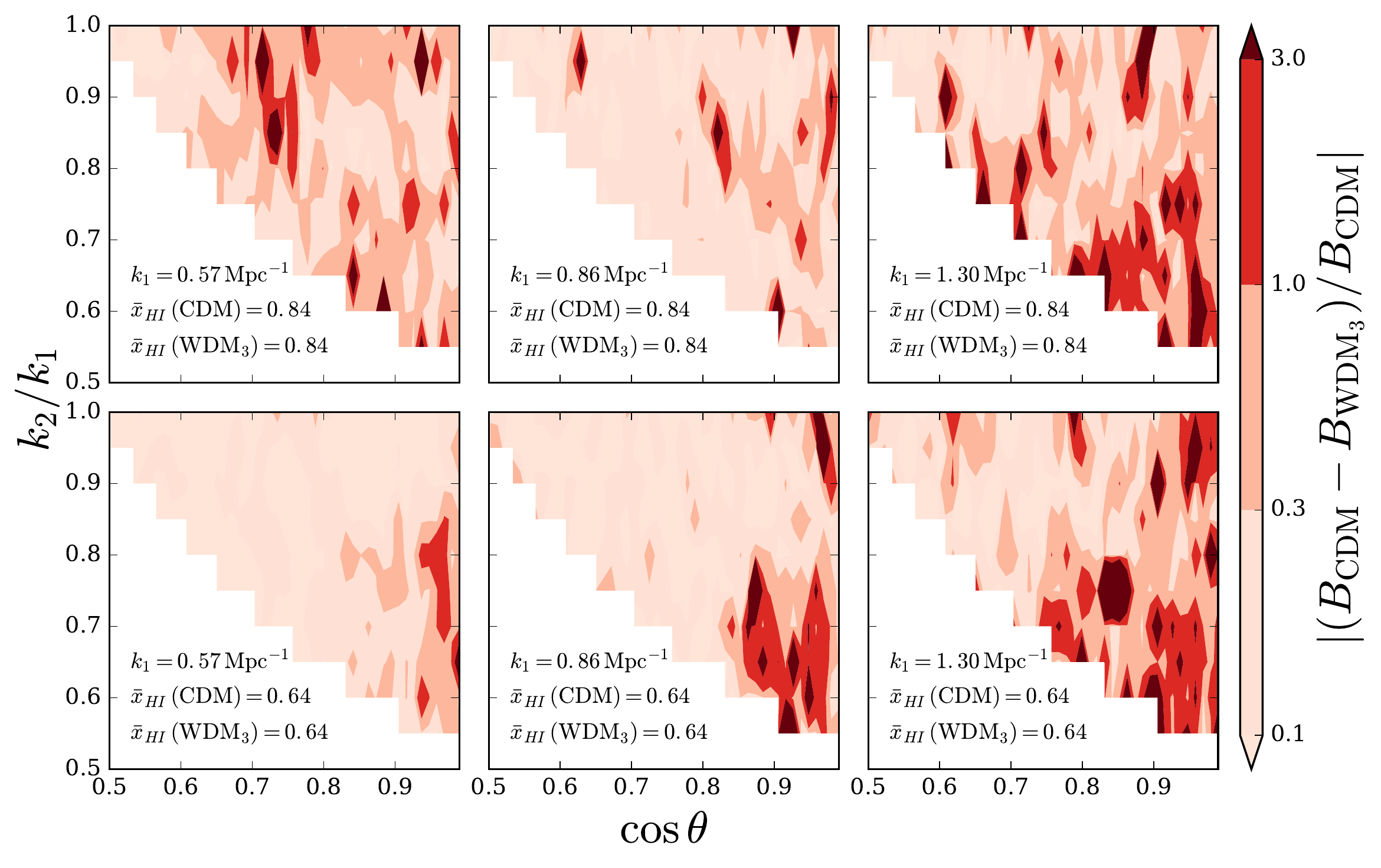}
    \caption{Relative fractional difference in 21-cm bispectrum between CDM and 2 keV WDM model (top panel), and between CDM and 3 keV WDM model (bottom panel) for $k_{1}$ = 0.57, 0.86 and 1.30 Mpc$^{-1}$ (left to right) at the same mass averaged neutral fraction with $\bar{x}_{\HI} = 0.84$ and $0.64$}
    \label{fig:diff_Bk}
\end{figure}
For the similar reasons as discussed in case of the power spectra, here also we compare the bispectra for different dark matter models approximately at the same stages of reionization (i.e. $\bar{x}_{\HI} = 0.84$ and $0.64$). We estimate the relative differences in the bispectra between the CDM and WDM models by computing the quantity $|(B_{\rm CDM} -B_{\rm WDM})/B_{\rm CDM}|$. We have shown this difference for isosceles and equilateral triangles in Figure~\ref{fig:bk_isosceles} and for all unique triangles in Figure~\ref{fig:diff_Bk}. It is apparent from these figures that at any stage of reionization, the differences between the $2$ keV WDM and CDM models is larger than the differences between the $3$ keV WDM and CDM models. For most of the $k$-triangle parameter space ($n-\cos{\theta}$ space) the relative difference between the $2$ keV WDM and CDM models is in between $30-300\%$ or more and for the $3$ keV WDM and CDM models the same  is in between $10-100\%$ or more. These differences are more prominent for triangles with larger $k_1$ modes compared to the triangles with smaller $k_1$ modes. They are particularly large in the region of the $n-\cos{\theta}$ space where $n\leq 0.75$ and $0.75 \leq \cos{\theta} \leq 1.0$.

One important point to note here is that, though we are looking at the 21-cm statistics at almost same stages of reionization in different dark matter models, as the redshifts are different the underlying halo characteristics and distribution will vary from model to model (which can be quantified by their halo mass functions). These differences in the halo mass, their numbers, and their spatial distribution will lead to a difference in 21-cm topology across the dark matter models, even when the overall level of ionization remains more or less the same in all of these cases. The 21-cm signal power spectrum is expected to be not very sensitive to these differences in topology. However, the signal bispectrum will be very sensitive to them as these differences in topology leads to a significant difference in the non-Gaussian characteristics of the signal. One can expect that these significantly larger relative differences in the signal bispectra (compared to their relative differences in power spectra), when the underlying dark matter model is different, will be possible to detect with the upcoming highly sensitive SKA. Using sophisticated parameter estimation techniques, while using the signal bispectra as the target statistic, one may even be able to constrain the nature of the dark matter from such future radio interferometric observations of the EoR.

\section{Summary} \label{summary}
In this article we have attempted to quantify the impact of the different kind of warm dark matter models (in comparison with the standard cold dark matter models) on the reionization process and related 21-cm observables such as the power spectrum and bispectrum. We have considered $2$ keV and $3$ keV thermal  WDM models in this context. Using GADGET 2.0 Nbody simulation we observed that the non-negligible free streaming of the dark matter particles in the warm dark matter scenarios suppresses the matter density perturbations on small scales (see Figure~\ref{fig:odf}). Further, this suppression was more prominent in the $2$ keV WDM scenario compared to the $3$ keV WDM model because the lighter the WDM particle, the larger the free streaming scale. We also observed that the effect of this suppressed structure formation gets reflected in the halo mass function of all the dark matter models (Figure~\ref{fig:hmf}).

Using a semi-numerical model for reionization, we further observed that due to this suppression of the low mass halos (which are the dominant sources of ionizing photons) the overall reionization of the universe gets delayed in the WDM scenarios compared to the CDM model. This delay is likewise larger in the $2$ keV WDM model compared to the $3$ keV WDM model, when one keeps the ionizing photon production efficiency of halos same in all dark matter models. The suppression of low mass halo additionally introduces a significant difference in 21-cm brightness temperature topology in case of WDM models compared to the CDM model.

Next, we have quantified the differences in the two observable statistics of the EoR 21-cm signal, power spectrum and bispectrum, for different dark matter models. We found that, if the statistics from different dark models are compared at the same redshifts, the power spectrum differs significantly for both small and large $k$ modes for different matter models. However, when they are compared at the same stage of reionization i.e. at same $\bar{x}_{\HI}$ (i.e. by countering the effect of delayed reionization history), the differences in the $P(k)$ become significantly small at all scales between the WDM and CDM models. These become undetectable in case of small $k$ modes as they fall within the sample variance limits.

We have, for the first time, quantified the impact of WDM models on the EoR 21-cm signal using the bispectrum. The bispectrum is expected to be more sensitive to the difference in the dark matter model as it is capable of capturing the non-Gaussian features in the signal to which power spectrum is not sensitive. The source of non-Gaussianity in the EoR 21-cm signal is the evolving \HI\ topology during this period, which gets significantly affected by the suppression of the low mass halos in case of the WDM models. We find that the relative differences between the 21-cm bispectra for the WDM and CDM models are larger than their relative differences in the 21-cm power spectrum when compared at the same redshifts. Even when compared at the same stages of the EoR, the relative differences between the EoR 21-cm bispectra for WDM and CDM models varies between $10\%-300\%$ for all unique $k$-triangles. This level of relative differences in 21-cm bispectra for most of the unique $k$-triangle parameter space ensures that one should be able distinguish between the different dark matter models using the future radio interferometric observations of the EoR and this may even help one to constrain the WDM model parameters. Through this analysis we have established that the redshifted 21-cm bispectra is a unique and much more sensitive statistics than the power spectrum for differentiating the impact of different models of dark matter on the reionization process and one may be able to constrain the nature of the dark matter using this statistic from future observations of the CD-EoR through the SKA.       

Note that in this work we have not taken into account the impact of spin temperature fluctuations on the EoR 21-cm signal, which may have a significant effect on the 21-cm bispectrum. We have considered only a single model for reionization. However, the 21-cm topology and the resulting non-Gaussianity in the signal may change significantly if we change our reionization model. All of these effects will affect the 21-cm bispectrum. However, even if we take into account all of these effects into our formalism for a more accurate model of the EoR 21-cm signal, we expect that the differences in different dark matter models will still be prominently visible in the 21-cm bispectra.    

\section*{Acknowledgements}
MV is supported by PD51-INFN INDARK and
grant agreement ASI-INAF n.2017-14-H.0. SM would like to thank ICTP, Trieste and SISSA, Trieste for their generous hospitality during the two academic visits in 2019. This work and the collaboration with MV was initiated and a major portion of the simulation and related data analysis was done during these visits. Simulations have been run on the Ulysses supercomputer at SISSA.

\bibliographystyle{mnras}
\bibliography{refer}

\appendix
\section{Differences in the observable statistics of the 21-cm signal with different ionizing efficiency for different models} \label{changednion}
In this appendix, we repeat our analysis by varying the ionizing efficiency parameter $N_{\rm ion}$ in different dark matter models to ensure that we obtain the same $\bar{x}_{\HI}$ for all of these models in the same redshifts. We choose the  $N_{\rm ion}$ values such that (Table~\ref{tab:Nions}) in all models universe becomes $50\%$ ionized by redshift $8$.

\subsection{21-cm topology}
\begin{figure}
    \centering
    \includegraphics[width = \columnwidth]{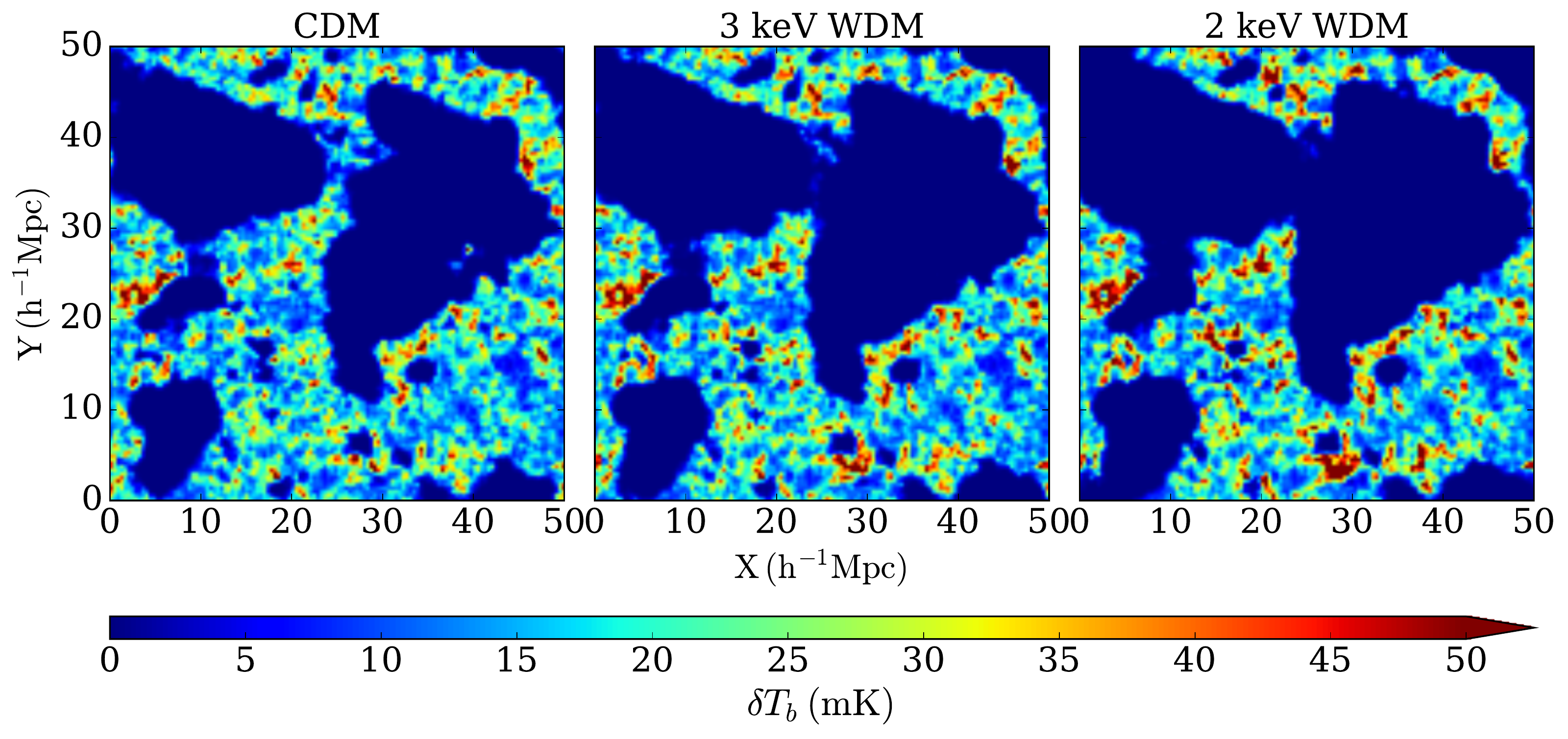}
    \caption{21-cm brightness temperature maps at $z = 8$ with mass averaged neutral fraction $\bar{x}_{\HI} = 0.47$. Left-hand panel: CDM; middle panel: 3 keV WDM; right-hand panel: 2 keV WDM.}
    \label{fig:diffNion_maps}
\end{figure}

In Figure~\ref{fig:diffNion_maps}, we have shown the 21-cm brightness temperature maps at $z = 8$ with mass averaged neutral fraction $\bar{x}_{\HI} \approx 0.5$ for all the models. We observe that the 21-cm topology has significant similarity at large scales. However, it has quite a few differences at small length scales for different models of dark matter. The size of ionized bubbles is relatively larger in WDM models compared to the CDM model. This is an obvious signature of the suppression in the number of low mass haloes in WDM models compared to the CDM model. The lack of low mass haloes have been compensated by increasing the ionizing photon production efficiency (see Table~\ref{tab:Nions}) in all haloes in these models. This implies that the high mass haloes in WDM models will produce significantly more photons compared to their counterparts in the CDM model and will thus produce larger ionized bubbles (compare the three panels in Figure~\ref{fig:diffNion_maps}). These features have also been reported by \citet{2018JCAP...08..045D}. 

\subsection{Observable statistics of the 21-cm signal}
\subsubsection{\textit{21-cm power spectrum}}
\begin{figure}
    \centering
    \includegraphics[width = \columnwidth]{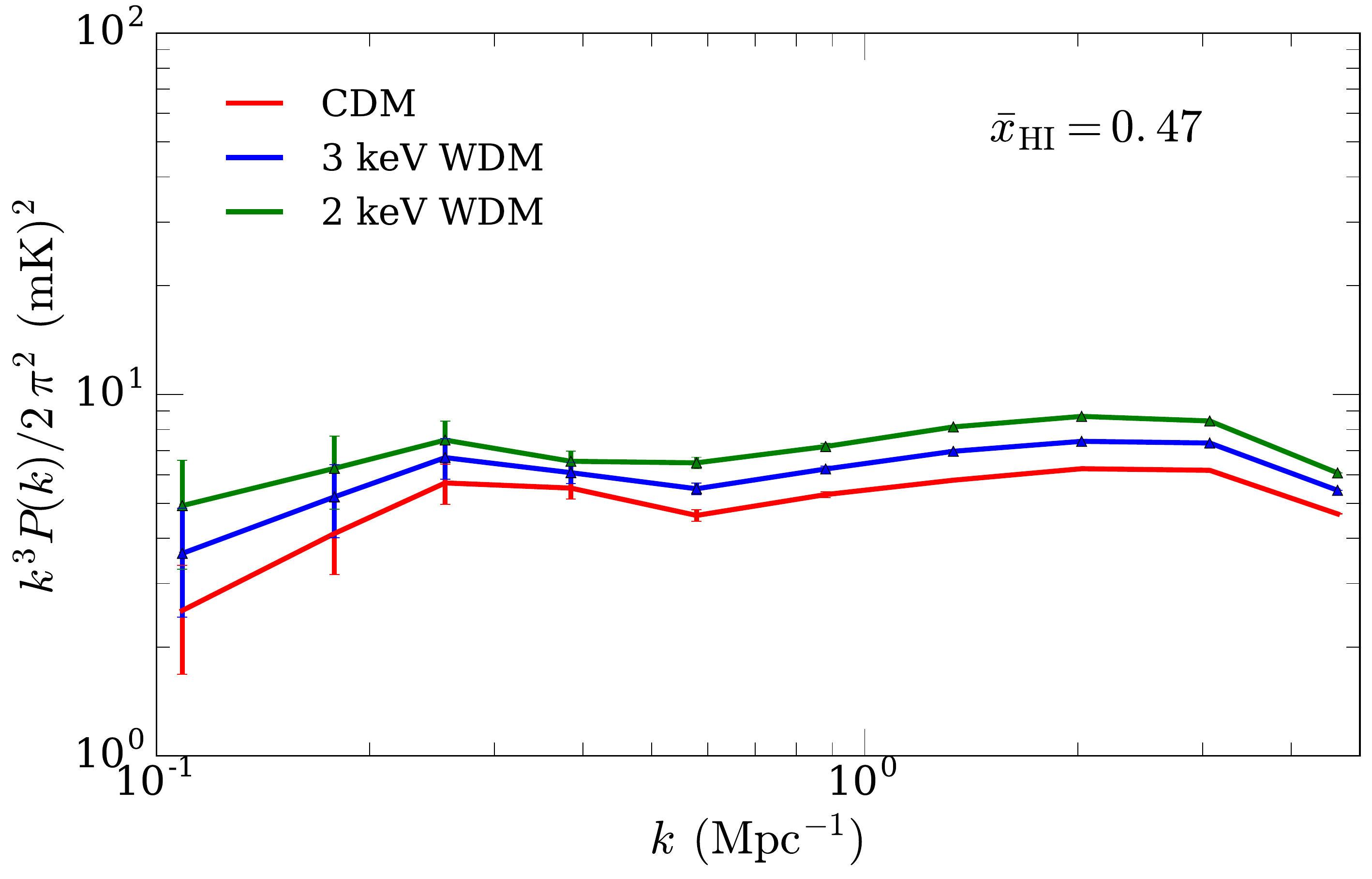}
    \caption{21-cm power spectrum at $z = 8$ with the mass averaged neutral fraction $\bar{x}_{\HI} = 0.47$ for CDM (red), 3 keV WDM (blue) and 2 keV WDM (green) model.}
    \label{fig:diffNionpk}
\end{figure}

Figure~\ref{fig:diffNionpk} shows the 21-cm power spectrum for different dark matter models at $\bar{x}_{\HI} \approx 0.5$. We observe that the 21-cm power spectrum in WDM models is greater than that in the CDM model. This is because in WDM models, the ionized bubbles are larger in size (due to the increased $N_{\rm ion}$ value). One can also notice that the difference in 21-cm power spectra between different models remains significantly low even if we change our reionization model, and at large scales these differences are within the sample variance limit.
\begin{table}
\centering
\caption{This tabulates the $N_{\rm ion}$ values required in different dark matter models to ensure that $\bar{x}_{\HI} \approx 0.5$ at $z = 8$.}
\label{tab:Nions}
\begin{tabular}{ccc}
\hline
CDM & 3 keV WDM & 2 keV WDM\\ [0.5ex]
\hline
\hline
23.21 & 33.70 & 47.50 \\
\hline
\end{tabular}
\end{table}

\subsubsection{\textit{21-cm bispectrum}}
\begin{figure}
    \centering
    \includegraphics[width = \columnwidth]{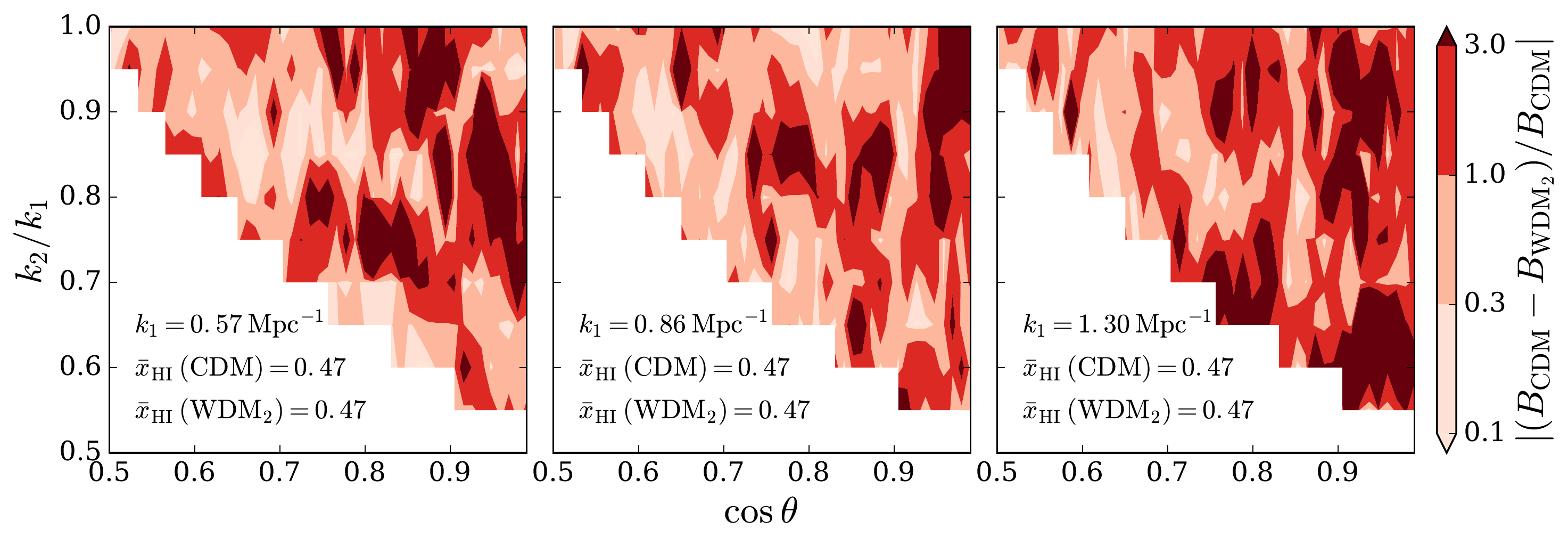}\\[0.2cm]
    \includegraphics[width = \columnwidth]{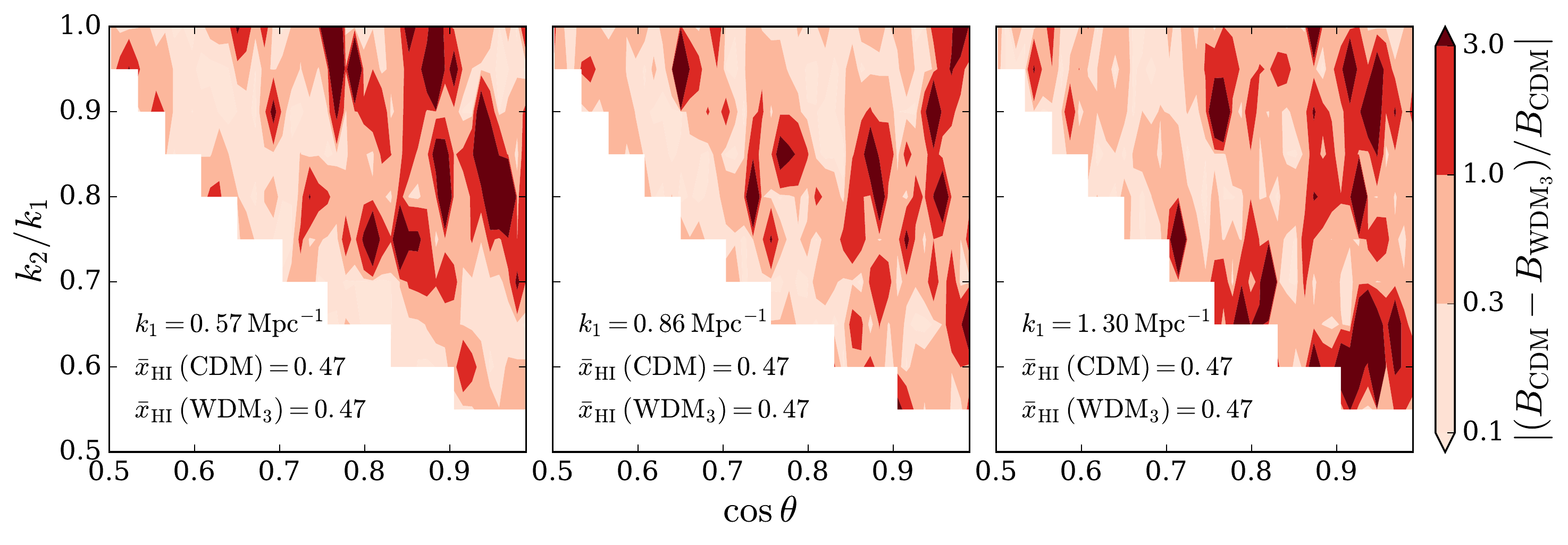}
    \caption{Relative fractional difference in 21-cm bispectra between CDM and 2 keV WDM model (top panel), and between CDM and 3 keV WDM model (bottom panel) at $k_{1} = 0.57, 0.86$ and $1.30$ Mpc$^{-1}$ (left to right) at mass averaged neutral fraction $\bar{x}_{\HI} = 0.47$}
    \label{fig:diffNionbk}
\end{figure}

In Figure~\ref{fig:diffNionbk}, we have shown the relative fractional difference in 21-cm bispectra between the CDM and WDM models at $k_{1} = 0.57,\, 0.86$ and $1.30$ Mpc$^{-1}$. One obvious observation that one can make is that these differences in bispectra are larger when estimated between CDM and 2 keV WDM model. 

Further, these differences are larger than those observed in Figure~\ref{fig:diff_Bk}. This is because by changing the $N_{\rm ion}$, we are changing our reionization model. Note that even at large scales $k_{1} = 0.57$ Mpc$^{-1}$, the differences are large for most of the unique $k$ triangles. Additionally, at relevant small scales (or large $k_{1}$ values), the differences are also quite large because the strength of the 21-cm signal gets increased in WDM models at small scales.

We conclude that even when one changes the reionization source model to arrive at the same state of IGM ionization at the same redshift for different dark matter models, the bispectrum remains equally or more sensitive to the characteristics of the dark matter model. However, the power spectrum remains equally insensitive to the dark matter model characteristics in this case as well.  

\bsp	
\label{lastpage}
\end{document}